\documentclass[prb,twocolumn,showpacs,floatfix]{revtex4}
\usepackage{graphicx}
\usepackage{amssymb}
\usepackage{amsmath}
\usepackage{xspace}
\usepackage{url}

\let \Im \relax
\DeclareMathOperator{\Im}{Im}
\newcommand{\phdag}{{\phantom{\dag}}}
\newcommand{\abs}[1]{\left\vert#1\right\vert}

\begin{document} 
\title{Many-body theory of the neutralization of strontium ions on gold surfaces}

\author{M. Pamperin, F. X. Bronold, and H. Fehske}
\affiliation{Institut f{\"ur} Physik,
             Ernst-Moritz-Arndt-Universit{\"a}t Greifswald,
             17489 Greifswald,
             Germany}
\date{\today}
\begin{abstract}
Motivated by experimental evidence for mixed-valence correlations affecting the neutralization 
of strontium ions on gold surfaces 
we set up an Anderson-Newns model for the Sr:Au system and calculate the neutralization
probability $\alpha$ as a function of temperature. We employ quantum-kinetic equations for the 
projectile Green functions in the finite$-U$ non-crossing approximation. Our results for $\alpha$ 
agree reasonably well with the experimental data as far as the overall order of magnitude is 
concerned showing in particular the correlation-induced enhancement of $\alpha$. The 
experimentally found non-monotonous temperature dependence, however, could not be reproduced.
Instead of an initially increasing and then decreasing $\alpha$ we find over the whole 
temperature range only a weak negative temperature dependence. It arises however clearly 
from a mixed-valence resonance in the projectile's spectral density and thus  
supports qualitatively the interpretation of the experimental data in terms of a 
mixed-valence scenario.
\end{abstract}


\maketitle

\section{Introduction}

Charge-transferring atom-surface collisions~\cite{Monreal14,Winter07,Winter02,Rabalais94,LG90,BN89,Modinos87,YM86,NMB83} 
are of great technological interest in surface science. The complex process
of neutral gas heating in fusion plasmas,~\cite{KFF07} for instance, starts
with the surface-based conversion of neutral hydrogen atoms to negatively charged ions. The 
operation modii of low-temperature plasmas used, for instance, in flat panel displays or in 
surface modification devices depend strongly on secondary electrons originating from the 
substrate due to impact of ions and radicals and thus also on surface-based charge-transfer 
processes.~\cite{LL05} Many surface diagnostics, finally, for instance, ion neutralization 
spectroscopy~\cite{Rabalais03} and meta-stable atom de-excitation 
spectroscopy~\cite{HMO97} utilize charge-transfer processes to 
gain information about the constituents of the surface. At the same time, however, 
charge-transferring atom-surface collisions are of fundamental interest as well because 
they are particular realizations of a quantum-impurity system out of equilibrium. 

The archetypical quantum-impurity is a local spin (more generally,
a local moment) in a metal coupled to the itinerant electrons of the conduction band. 
Its well-documented properties,~\cite{Hewson93,Fulde95} arising from an emerging 
resonance at the Fermi energy of the metal, are however also present in other quantum 
systems with a finite number of correlated internal states interacting via tunneling with 
a reservoir of external states. In particular, semiconductor quantum-dots coupled to 
metallic leads are ideal platforms for studying local-moment physics in a well-controlled 
setting.~\cite{CC09,PG04,AL03,GSM98,COK98,GGK98,WM94,GD92} By a suitable
time-dependent gating the dot can be driven out of equilibrium.
Of particular recent theoretical interest are the temporal build-up 
and/or decay of local-moment-type correlations and how they affect the electron transport 
through these devices.~\cite{NC14,LA14,MUK11,CR11} As pointed out a long time ago by Shao and 
coworkers~\cite{SNL96} as well as Merino and Marston~\cite{MM98}, similar transient correlations 
should also occur in charge-transferring atom-surface collisions where the projectile with its 
finite number of electron states mimics the quantum dot while the target with its continuum 
of states replaces the lead. 

A recent experiment by He and Yarmoff indeed provided strong evidence for local-moment-type 
correlations to affect the neutralization probability of strontium ions on gold 
surfaces.~\cite{HY11,HY10} 
They found a non-monotonous temperature dependence of the neutralization probability which 
first increases and then decreases with temperature. The initial increase with temperature 
is most probably a thermal single-particle effect but the latter could be the long-sought 
fingerprint for a transient mixed-valence resonance formed during an
electron-transfer from a surface to an atomic projectile.~\cite{SNL96,MM98}

In the present work, following the lead of Nordlander and coworkers~\cite{SNL96,SLN94a,SLN94b,LN91} 
as well as Merino and Marston,~\cite{MM98} we analyze He and Yarmoff's experiment~\cite{HY11,HY10} 
from a many-body theoretical point of view. In particular we test the claim that the 
negative temperature dependence at high temperatures arises from the local moment of the unpaired 
electron in the $5s$ shell of the approaching ion. For that purpose we first set up, as usual 
for the description of charge-transferring atom-surface collisions, an 
Anderson-Newns Hamiltonian.~\cite{LG90,BN89,Modinos87,YM86,KO87,NKO88,NMB83,RFG09,BFG07,GFM05,OM96,MAB93}
To obtain its single-particle matrix elements we employ Hartree-Fock wave functions for the 
strontium projectile,~\cite{CR74} a step-potential description for the gold target, and Gadzuk's 
semi-empirical construction~\cite{Gadzuk67a,Gadzuk67b,Gadzuk09} for the 
projectile-target interaction. The model rewritten in terms of Coleman's pseudo-particle 
operators~\cite{Coleman84,KR86} 
is then analyzed within the finite$-U$ non-crossing approximation employing contour-ordered 
Green functions~\cite{KB62,Keldysh65} as originally suggested by 
Nordlander and coworkers.~\cite{SNL96,SLN94a,SLN94b,LN91} Besides the instantaneous 
occupancies and the neutralization probability we also calculate the instantaneous spectral 
densities. The latter are of particular interest because if the interpretation of the 
experimental findings in terms of a mixed-valence scenario is correct, the projectile's 
spectral density should feature a transient resonance at the target's Fermi energy.

For the material parameters best suited for the Sr:Au system we
find neutralization probabilities slightly above the experimental data but still of the 
correct order of magnitude indicating that the single-particle matrix elements of the 
Anderson-Newns model are sufficiently close to reality. Moreover, for the model without 
correlations the neutralization probabilities turn out to be too small showing 
that agreement with experiment can be only achieved due to the correlation-induced 
enhancement of the neutralization probability. We also find a transient resonance in 
the instantaneous spectral densities hinting mixed-valence correlations to be present 
in certain parts of the collision trajectory. The non-monotonous temperature dependence 
of the neutralization probability, however, could not be reproduced. Instead we find 
the resonance to lead only to a weak negative temperature dependence over the whole 
temperature range. 

Due to lack of data for comparison we cannot judge the validity of the single-particle 
parameterization we developed for the Sr:Au system. At the moment it is the most 
realistic one. We attribute therefore the failure of the present calculation to reproduce 
the temperature anomaly of the neutralization probability while having at the same
time mixed-valence features in the instantaneous spectral densities primarily to the 
finite$-U$ non-crossing approximation which seems to be unable to capture the instantaneous 
energy scales with the required precision. A quantitative description of the experiment has 
thus to be based either on the dynamical $1/N$ expansion initially used by Merino and 
Marston,~\cite{MM98} equation of motions for the correlation functions of the physical
degrees of freedom instead of the pseudo-particles,~\cite{GFM05} or on the one-crossing 
approximation as it has been developed for the equilibrium Kondo 
effect.~\cite{OK06,KW05,PG89,HS89,Sakai88} 
Numerically this will be rather demanding. But demonstrating that He and Yarmoff 
have--for the first time--indeed seen local-moment physics in a charge-transferring atom-surface
collision may well be worth the effort.

The paper is organized as follows. In the next section we introduce the Anderson-Newns 
model, its parameterization for the Sr:Au system, and its representation in terms of 
pseudo-particle operators. In Sec.~\ref{QuantumKinetics} we recapitulate briefly the 
quantum kinetics of the Anderson-Newns model as pioneered by Nordlander and coworkers. 
Basic definitions and the main steps of the derivation of the 
set of Dyson equations for the analytic pieces of the projectile Green functions within 
the finite-$U$ non-crossing approximation, which is the set of equations to be numerically 
solved, can be found in an appendix to make the paper self-contained. 
Numerical results are presented, discussed, and compared to the experimental 
data in Sec.~\ref{Results} and concluding remarks with an outlook are given in Sec.~\ref{Conclusions}.

\section{Model}
\label{Model}

The interaction of an atomic projectile with a surface is a complicated many-body process. 
Within the adiabatic approximation, which treats the center-of-mass motion of the projectile 
along the collision trajectory classically,~\cite{YM86} it leads to a position- and hence 
time-dependent broadening and shifting of the projectile's energy levels. The adiabatic 
modification of the atomic energy levels as a function of distance can be calculated from first 
principles.~\cite{NT88,NT90,BW95,MMM98,VGB05} As in our previous work on secondary electron 
emission from metallic~\cite{MBF12a} and dielectric~\cite{MBF12b,MBF11} surfaces, we employ however 
Gadzuk's semi-empirical approach~\cite{Gadzuk67a,Gadzuk67b}--based on classical image shifts and a 
golden rule calculation of the level widths--which not only provides a very appealing physical 
picture of the interaction process~\cite{Gadzuk09} but produces for distances larger than a few 
Bohr radii also reasonable level widths and shifts.~\cite{BW95,MMM98,VGB05} 

Indeed, first-principle investigations of Auger neutralization of helium ions on aluminum surfaces 
by Monreal and coworkers~\cite{MMM98,VGB05} showed that for distances larger than five Bohr radii 
the level shift follows the classical image shift. Only for shorter distances chemical interactions
lead to a substantial deviation between the two. Borisov and Wille,~\cite{BW95} on the other hand,
found the level width of hydrogen ions approaching an aluminum surface to be for distances larger 
than five Bohr radii also not too far off the widths obtained from Gadzuk's golden rule calculation, 
that is, the widths are perhaps off by a factor two. The reason most probably is Gadzuk's ingenious 
choice of the tunneling matrix element (see below) which takes care of the non-orthogonality
of the projectile and target states.~\cite{Gadzuk67b} Since the turning point of the strontium ion is 
sufficiently far away from the first atomic layer, we estimate it to be around five Bohr 
radii, we expect Gadzuk's semi-empirical approach to also provide a reasonable parameterization 
of the Sr:Au system investigated by He and Yarmoff.~\cite{HY10,HY11} 
The corrections due to chemical interactions between the strontium projectile and the gold 
surface, occurring at shorter distances and included in first-principle 
approaches,~\cite{NT88,NT90,BW95,MMM98,VGB05} should not yet play a role.

\begin{figure}[t]
\includegraphics[width=0.9\linewidth]{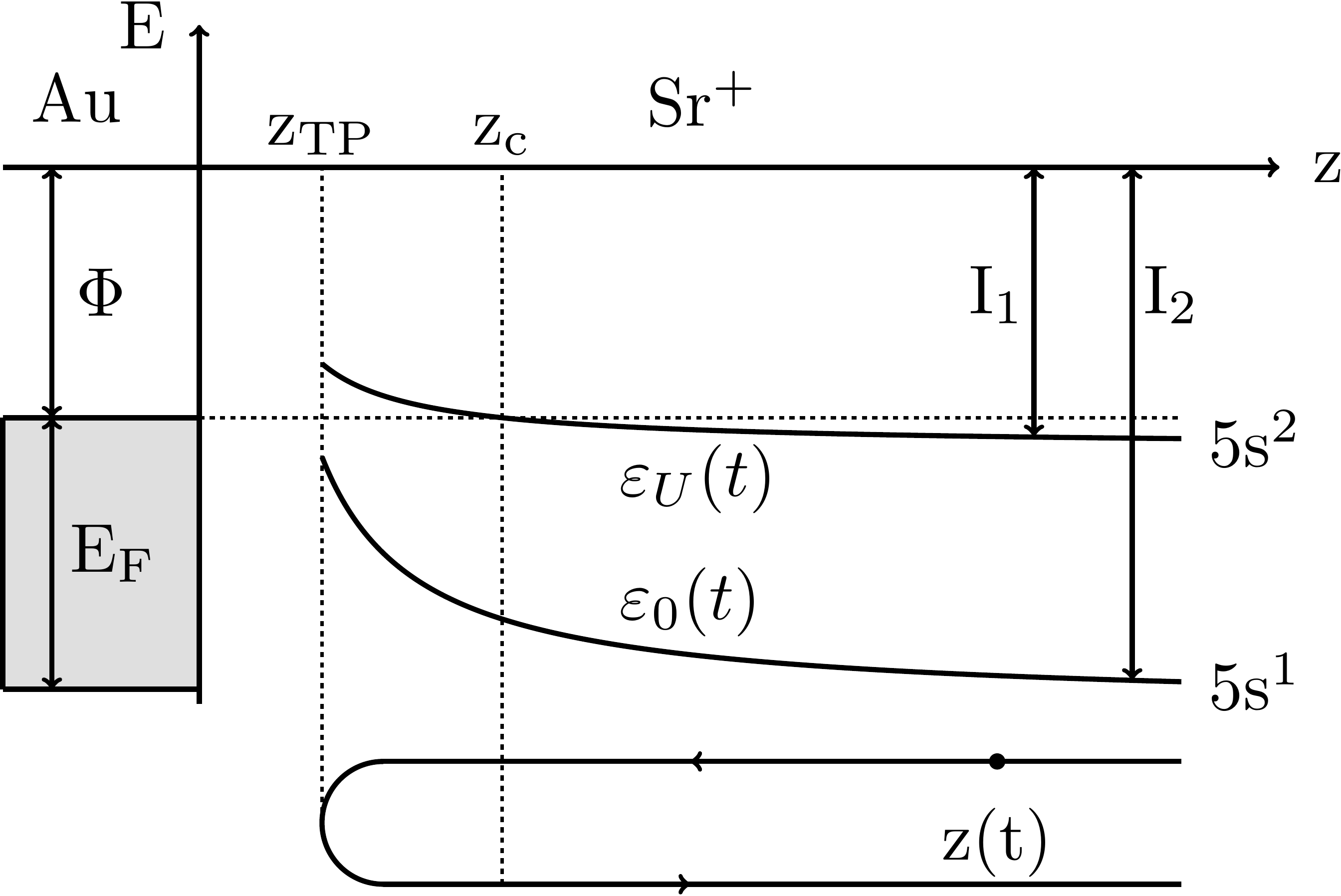}
\caption{Illustration of the time-dependent charge-transferring scattering of a 
$\mathrm{Sr^+}$ ion on a gold surface. Far from the surface the two ionization energies
$\varepsilon_0 (t)$ and $\varepsilon_U (t)$ are equal to the first two ionization energies of 
a strontium atom and represent the projectile's $\mathrm{5s^1}$ and $\mathrm{5s^2}$ configuration. 
They shift upward due to the image interaction of the projectile with the surface. For 
simplicity not shown is the hybridization between the projectile and surface states as
the projectile closes its distance to the surface which is idealized by a step-potential whose
depth is the sum of the work function $\Phi >0$ and the Fermi energy $E_F >0$. The energies are 
on scale and the points indicated along the trajectory $z(t)$ are $z_{\rm TP}$, the turning point, 
and $z_c$, the point where the first ionization level crosses the Fermi energy.}
\label{schematic}
\end{figure}

We now set up Gadzuk's approach step by step.~\cite{Gadzuk67a,Gadzuk67b}
For the charge-transfer process we are interested in, the first two ionization levels of the 
strontium projectile are most important. They are closest to the Fermi energy of the gold 
target and may hence accept or donate an electron. In terms of the Anderson-Newns model,
the two levels constitute, respectively, the upper and lower charge-transfer level. 
The difference of the two can thus be identified with the time-dependent on-site Coulomb 
repulsion. Figure~\ref{schematic} schematically shows the essence of the Anderson-Newns model 
for the Sr:Au system. The energy levels are on scale. Mathematically, the on-site energies are given by 
\begin{align}
\varepsilon_U (t) &= -I_1 + \dfrac{e^2}{4\abs{z(t)-z_i}}~, \\
\varepsilon_0 (t) &= -I_2 + \dfrac{3e^2}{4\abs{z(t)-z_i}}~,
\end{align}
where $I_1>0$ and $I_2>0$ are the first and second ionization energies far from the 
surface and $z_i$ is the distance of the metal's image plane from its crystallographic ending 
at $z=0$. For simplicity the projectile is assumed to approach the surface perpendicularly 
along the trajectory,
\begin{align}
z(t) = z_{\rm TP} + v \abs{t}~,
\end{align}
where $z_{\rm TP}$ is the turning point and $v$ is the velocity.

The shift of the on-site energies with time can be interpreted as the negative of the energy gain 
of a virtual process which moves the configuration under consideration from the actual position 
$z$ to $z=\infty$, reduces its electron occupancy by one, and then moves it back to its former 
position $z$, taking into account in both moves possible image interactions due to the initial 
and final charge-state of the projectile with the metal.~\cite{NMB83} 

For the upper level, 
$\varepsilon_U(z)$, corresponding to the first ionization level, this means shifting the charge-neutral
$5s^2$ configuration from $z$ to $z=\infty$, turning it into a single-charged $5s^1$ configuration, 
which is then moved back to $z$. In the first leg no image shift occurs while in the second one
the image shift is $-e^2/4|z-z_i|$. The net energy gain of the whole process is therefore 
$-e^2/4|z-z_i|$ leading to a shift of the upper on-site level of $+e^2/4|z-z_i|$. Similarly, for the 
lower level, $\varepsilon_0(z)$, which is the second ionization level, one imagines moving a $5s^1$ 
configuration from $z$ to $z=\infty$ and then a $5s^0$ configuration from $z=\infty$ back to $z$. 
In both moves image shifts occur adding up to $-3e^2/4|z-z_i|$ because the energy pay in the first 
half of the trip is due to a single-charged projectile while the energy gain on the return trip 
arises from a double-charged one. The shift of the lower on-site level is thus $+3e^2/4|z-z_i|$.
\begin{table}[t]
\begin{center}
  \begin{tabular}{c|c|c|c|c|c|c|c|c}
           & $I_1$[eV] & $Z_1$ & $I_2$[eV] & $Z_2$ & $\Phi$[eV] & $E_F$[eV] & $z_{\rm i}$[a.u.] & $m_e^*/m_e$ \\\hline
  {\rm Sr} & 5.7 & 1.65 & 11.0 & 2 & -- & -- & -- & -- \\
  {\rm Au} & -- & -- & -- & -- & 5.1 & 5.53 & 1.0 & 1.1 \\
  \end{tabular}
  \caption{Material parameters for strontium and gold: $I_1$ and $I_2$ are the
  first and the second ionization energy, $Z_1$ and $Z_2$ are the effective charges to be
  used in the calculation of the hybridization matrix element (viz: Eq.~\eqref{TME}),
  $\Phi$ is the work function, $E_F$ the Fermi energy, $z_i$ the position of the image 
  plane in front of the surface for which we take a typical value, and $m_e^*$ is the effective 
  mass of an electron in the conduction band of gold.}
  \label{Parameters}
\end{center}
\end{table}
 
Besides the on-site energies we also need the hybridization matrix elements which depend 
on projectile and metal wave functions. Ignoring the lateral variation of the 
surface potential, we take for the latter simply the wave functions of a step potential 
with depth $V_0 = -\Phi - E_F$ where $\Phi > 0$ and $E_F>0$ are the work function and the 
Fermi energy of the surface, respectively, measured as illustrated in Fig. \ref{schematic}. 
Hence, the energies and wave functions for the conduction band electrons are
\begin{align}
\varepsilon_{\vec{k}} & = \frac{\hbar^2}{2 m^*_e} \left( k_x^2 + k_y^2 + k_z^2 \right)-\abs{V_0}~,\\
\psi_{\vec{k}} (\vec{r})& = \dfrac{1}{L^{3/2}} e^{i(k_x x + k_y y)} \Bigl\{ \Theta(z) T_{k_z} e^{-\kappa_{z} z} 
\nonumber \\
&+ \Theta(-z) \left[ e^{i k_z z} + R_{k_z} e^{-i k_z z} \right] \Bigr\}~,
\end{align}
where $L$ is the spatial width of the step, which drops out in the final expressions, and  
\begin{align}
T_{k_z} = \frac{2 i k_z}{i k_z - \kappa_{z}}~,~~
R_{k_z} = \frac{i k_z + \kappa_{z}}{i k_z - \kappa_{z}}~,
\end{align}
with $\kappa_{z} = \sqrt{2 m_e(\abs{V_0}-k_z^2)/\hbar^2}$ are, respectively, the 
transmission and reflection coefficients of the step potential. More sophisticated 
surface potentials are conceivable but from the work of K\"urpick and Thumm~\cite{KT96}
we expect the final result for the neutralization probability to depend not too 
strongly on the choice of the surface potential.

For the calculation of the hybridization matrix element we also need $5s$ wave functions for 
the neutral and single-charged projectile. Both are radially symmetric and in the 
Hartree-Fock approximation can be written in the general form
\begin{align}
\psi_{\rm{HF}} (\vec{r}\,) = \sum_{j=1}^N \frac{c_j N_j}{\sqrt{4\pi}} \abs{\vec{r}\,}^{n_j-1} 
e^{-C_j \abs{\vec{r}\,}}
\end{align}
with $N$, $c_j$, $N_j$, $n_j$ and $C_j$ tabulated parameters.~\cite{CR74}

The transfer of an electron between the target and the projectile is a re-arrangement collision. 
According to Gadzuk~\cite{Gadzuk67a,Gadzuk67b} the matrix element for this process which is also
the hybridization matrix element of the Anderson-Newns model is given by 
\begin{align}
V_{\vec{k}} (t) = \int_{z>0} \!\! d^3 r \, \psi_{\vec{k}}^{*} (\vec{r}\,) \frac{Ze^2}{|\vec{r}-\vec{r}_p(t)|} 
\psi_{\rm{HF}} ({\vec{r}-\vec{r}_p (t)})~,
\label{TME}
\end{align}
where the potential between the two wave functions is the Coulomb interaction of the transferring
electron with the core of the projectile located at $\vec{r}_p(t)=z(t)\vec{e}_z$. This choice 
of the matrix element takes into account the non-orthogonality of the projectile and target 
states.~\cite{Gadzuk67b} The charge of 
the core, $Ze$, is screened by the residual valence electrons of the projectile, that is, 
for the lower level $Z=Z_2=2$ while for the upper level $Z=Z_1=2-s$ with $s = 0.35$ the Slater 
shielding constant for a $5s$ electron.~\cite{Slater30} Material parameters required for the
modeling of the Sr:Au system are listed in Table~\ref{Parameters}.

The multidimensional integral~(\ref{TME}) can be analytically reduced to a one-dimensional 
integral by a lateral Fourier transformation of the product of the residual Coulomb
interaction with the Hartree-Fock projectile wave function. The resulting sum contains 
modified Bessel functions of the second kind $K_\alpha$.~\cite{AS73} Transforming formally 
back and reversing the order of integration yields after successively integrating first along 
the $x$, $y$ and then along the $p_x$, $p_y$ directions  
\begin{widetext}
\begin{align}
V_{\vec{k}} (t) &= \sqrt{2} \dfrac{Ze^2}{L^{3/2}} T_{k_z}^* (\theta, \phi)  \sum_{j=1}^{N} 
\sum_{n=0}^{(n_j-1)/2} (-1)^n c_j N_j C_j^{n_j-1-2n}  
\Bigl( C_j^2 + k_x^2 + k_y^2 \Bigr)^{-(n_j-1/2-n)/2}  A_{n_j n} \nonumber \\
&\times \int_{0}^\infty \!\! dz \, e^{-\kappa_{z} z} \, \abs{z-z_p(t)}^{n_j-1/2-n} K_{n_j-1/2-n} 
\Bigl(\abs{z-z_p(t)} \sqrt{C_j^2 + k_x^2 + k_y^2} \Bigr)~,
\label{Vkfinal}
\end{align}
\end{widetext}
where 
\begin{align}
A_{n_j n} = \begin{pmatrix}
                                1 & 0 & 0 \\
                                1 & 0 & 0 \\
                                1 & 1 & 0 \\
                                1 & 3 & 0 \\
                                1 & 6 & 3
                        \end{pmatrix}~.
\end{align}
are numerical coefficients ($n_j\in\{1,...,5\}$ and $n=0,1,2$ for $5s$ functions~\cite{CR74})
and $T_{k_z}^*$ is the complex conjugate of $T_{k_z}$. 

Inserting the matrix element~\eqref{Vkfinal} into the golden rule expression for the transition rate 
gives the level width
\begin{align}
\Gamma_{\varepsilon (t)} (t) = \frac{2\pi}{\hbar} 
\sum_{\vec{k}} \left|V_{\vec{k}} (t)\right|^2 \delta(\varepsilon (t)
-\varepsilon_{\vec{k}})~.
\label{GammaEq}
\end{align}
It is an important quantity characterizing the strength of the charge-transfer. Turning the 
momentum summation into an integral eliminates the width $L$ of the step potential. The integrals 
have to be done numerically and lead due primarily to the modified Bessel functions to level widths 
exponentially decreasing with distance as it is generally expected.

\begin{figure}[t]
\includegraphics[width=0.9\linewidth]{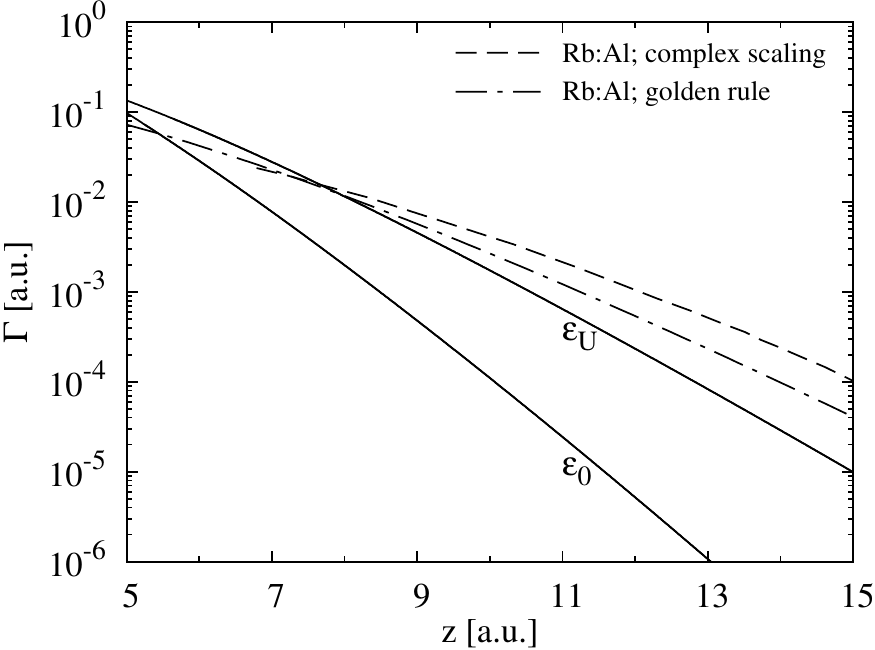}
\caption{Widths of the first ($\varepsilon_U$) and second ($\varepsilon_0$) ionization level 
of a strontium projectile approaching a gold surface as computed from Eq.~(\ref{GammaEq}) on
the basis of Hartree-Fock wave functions for the projectile and simple step-potential wave
functions for the target. Atomic units are used, that is, energy is measured in Hartrees
and length in Bohr radii. Data are shown only for distances larger than 5 Bohr radii,
which is the turning point of the collision trajectory. The width of a rubidium 5s level 
is also shown and contrasted with the width obtained for that level by Nordlander 
and Tully using the complex-scaling approach.~\cite{NT90} Notice, in contrast to 
Nordlander and Tully's Eq.~(4.1)~\cite{NT90} our widths~\eqref{GammaEq} contain a factor 
$2\pi$ and not a factor $\pi$. For the comparison we corrected for this difference.}
\label{GammaPic}
\end{figure}

In Fig.~\ref{GammaPic} we show the widths of the first two ionization levels of the strontium 
projectile hitting a gold surface as obtained from Eq.~(\ref{GammaEq}) by setting $\varepsilon (t)$ 
to $\varepsilon_U (t)$ and $\varepsilon_0 (t)$, respectively, and using the material parameters 
given in Table~\ref{Parameters}. To demonstrate that the widths we get are of the correct 
order of magnitude, we also plot the width of a rubidium 5s level in front of an aluminum surface
and compare it with the width obtained by Nordlander and Tully using a complex scaling technique.~\cite{NT90}
In qualitative agreement with Borisov and Wille's investigation~\cite{BW95} of Gadzuk's approach 
our rubidium width is a factor 2-3 too small for $z > 7 a_B$ and a factor 2 too large for 
$z = 5 a_B$. Between $7 a_B$ and $9 a_B$, however, the widths fortuitously agree with each other. 
The same trend we found for the other alkaline-metal combinations investigated by Nordlander 
and Tully.~\cite{NT90} From this comparison we expect the widths of the strontium levels to be 
of the correct order of magnitude for intermediate distances between $5 a_B$ and $12 a_B$. This 
is the range required for the description of the collision process we are interested in. For smaller 
and larger distances the semi-empirical approach breaks down and should be replaced by 
quantum-chemical methods.~\cite{NT88,NT90,BW95,MMM98,VGB05} 

\begin{figure}[b]
\includegraphics[width=0.75\linewidth]{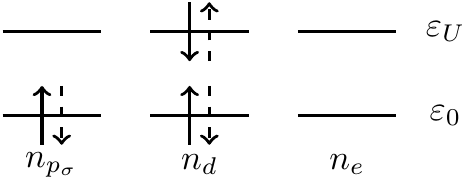}
\caption{Coleman's pseudo-particle representation for the strontium projectile. Shown are
the occupancies of the two ionization levels $\varepsilon_0$ and $\varepsilon_U$. The two
single-occupied configurations $n_{p_\sigma}$ contain an electron only in the second
ionization level. In the double-occupied state $n_d$ both ionization levels are occupied
by electrons with opposite spin whereas in the empty configuration
non of the ionization levels is occupied.}
\label{pseudostatesPic}
\end{figure}

With the single-particle matrix elements at hand the 
Anderson-Newns Hamiltonian~\cite{LG90,BN89,Modinos87,YM86,KO87,NKO88,RFG09,BFG07,GFM05,OM96,MAB93} 
describing the charge-transfer between the strontium ion and the gold 
surface is given by 
\begin{align}
H(t) &= \sum_\sigma \varepsilon_0(t) c_\sigma^\dagger c_\sigma^\phdag + 
\sum_{\vec{k} \sigma} \varepsilon_{\vec{k}} c_{\vec{k} \sigma}^\dagger c_{\vec{k} \sigma}^\phdag
\nonumber\\
&+ \frac{1}{2}(\varepsilon_U(t)-\varepsilon_0(t)) \sum_\sigma c_\sigma^\dagger c_\sigma^\phdag c_{-\sigma}^\dagger c_{-\sigma}^\phdag \nonumber\\ 
&+ \sum_{\vec{k} \sigma} \big[ V_{\vec{k}} (t) c_{\vec{k} \sigma}^\dagger c_\sigma^\phdag + h.c. \big]~
\end{align}
with $c_\sigma^\dagger$ creating an electron with spin polarization $\sigma$ in the $5s$ shell of 
strontium and $c_{\vec{k}\sigma}^\dagger$ creating an electron with spin polarization $\sigma$ 
and momentum $\vec{k}$ in the conduction band of the gold surface. Using Coleman's pseudo-particle 
representation~\cite{Coleman84,KR86}
\begin{align}
c_\sigma^\phdag &= e^\dagger p_\sigma^\phdag + p_{-\sigma}^\dagger d^\phdag~, \label{PPRa}\\
c_\sigma^\dagger &= e^\phdag p_\sigma^\dagger + p_{-\sigma}^\phdag d^\dagger~
\label{PPRb}
\end{align}
with $e^\dagger$, $p_\sigma^\dagger$ and $d^\dagger$ creating, respectively, 
an empty (${\rm Sr}^{2+}$), a single-occupied (${\rm Sr}^{+}$), and a double-occupied 
(${\rm Sr}^0$) strontium projectile (see Fig.~\ref{pseudostatesPic}), the Hamiltonian 
becomes~\cite{SLN94b}
\begin{align}
H(t) &= \sum_\sigma \varepsilon_0 (t )p_\sigma^\dagger p_\sigma^\phdag + [\varepsilon_0 (t) 
+ \varepsilon_U (t)] d^\dagger d^\phdag \nonumber\\ &+ \sum_{\vec{k} \sigma} \varepsilon_{\vec{k}} 
c_{\vec{k} \sigma}^\dagger c_{\vec{k} \sigma}^\phdag + \sum_{\vec{k} \sigma} \big[ V_{\vec{k}} (t) 
c_{\vec{k} \sigma}^\dagger e^\dagger p_\sigma^\phdag + h.c. \big] \nonumber\\
&+ \sum_{\vec{k} \sigma} \big[ V_{\vec{k}} (t) c_{\vec{k} \sigma}^\dagger d^\phdag p_{-\sigma}^\dagger + h.c. \big]~,
\end{align}
where the pseudo-particle operators obey the constraint 
\begin{align}
Q = \sum_\sigma p_\sigma^\dagger p_\sigma^\phdag + d^\dagger d^\phdag + e^\dagger e^\phdag = 1
\end{align}
since only one of the four possible projectile configurations can be ever realized.

\section{Quantum kinetics}
\label{QuantumKinetics}

To calculate the probability for the neutralization of a strontium ion on a gold surface we employ 
the formalism developed by Nordlander and coworkers. The formalism, based on contour-ordered 
Green functions~\cite{KB62,Keldysh65}, has been developed in a series of papers.~\cite{SNL96,SLN94a,SLN94b,LN91} 
However, the finite$-U$ equations, which we have to adopt and solve for the Sr:Au system, can be 
found only in the book edited by Rabalais~\cite{SLN94b} which may no longer by easily accessible. It is thus 
helpful to briefly summarize the finite$-U$ quantum kinetics as it is applied to the problem at hand.
Basic definitions and the main steps of the derivation of the most relevant equations can be found 
in the appendix. 

The central objects of the formalism are the contour-ordered Green functions for the empty, single-, 
and double-occupied projectile. They are denoted, respectively, by 
$E(t,t^\prime)$, $P_\sigma (t,t^\prime)$, and $D(t,t^\prime)$. The analytic pieces of these 
functions can be factorized ($\hbar=1$) 
%
\begin{align}
H^{\rm{R}} (t,t^\prime) &= -i \Theta (t-t^\prime) \exp\left( -i \int_{t^\prime}^t \!\! d\bar{t} \, 
\varepsilon (\bar{t}\,)\right) \bar{H}^{\rm{R}}(t,t^\prime)~,\label{factorR}\\
H^{\gtrless} (t,t^\prime) &= \exp\left( -i \int_{t^\prime}^t \!\! d\bar{t} \, 
\varepsilon(\bar{t}\,)\right) \bar{H}^{\gtrless} (t,t^\prime)~,
\label{factorLess}
\end{align}
where $H(t,t^\prime)$ can be any of the three Green functions and $\varepsilon (t)$ is either identical 
to zero, $\varepsilon_0 (t)$ or $\varepsilon_0 (t) + \varepsilon_U (t)$, depending on the function. 
The superscripts, $R$, $<$, and $>$ stand for, respectively, retarded, less-than, and greater-than 
Green functions.

\begin{figure}[t]
\includegraphics[width=0.9\linewidth]{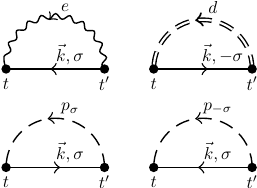}
\caption{Self-energies in the non-crossing approximation. Wavy, dashed, and double-dashed 
lines denote, respectively, fully dressed propagators for the empty (e), the  
single-occupied (p), and the double-occupied (d) configuration. The solid line is the bare 
Green function for the electrons of the surface. Starting at the left upper corner and 
proceeding clockwise the diagrams denote, respectively, the self-energies $\Sigma_{0,\sigma}$
$\Sigma_{U,\sigma}$, $\Pi_{d}$, and $\Pi_{e}$  for the Green function $P_\sigma$, 
$D$, and $E$.}
\label{selfPic}
\end{figure}

Using this notation and the non-crossing self-energies diagrammatically shown in Fig.~\ref{selfPic} 
gives after a projection to the $Q=1$ subspace~\cite{LN91,WM94,AL03} and an application of the 
Langreth-Wilkins rules~\cite{LW72} the equations of motion for the analytic pieces of the Green 
functions:
\begin{align}
\frac{\partial}{\partial t} \bar{E}^{\rm{R}} (t,t^\prime) = 
&-\sum_\sigma \int_{t^\prime}^t \!\! d\bar{t} \, \bar{K}_{\varepsilon_0}^{<} (\bar{t},t) 
\bar{P}_\sigma^{\rm{R}} (t,\bar{t}\,) \bar{E}^{\rm{R}} (\bar{t},t^\prime)~,
\label{Er}\\
\frac{\partial}{\partial t} \bar{P}_\sigma^{\rm{R}} (t,t^\prime) = 
&-\int_{t^\prime}^t \!\! d\bar{t} \, \bar{K}_{\varepsilon_0}^{>} (t,\bar{t}\,) 
\bar{E}^{\rm{R}} (t,\bar{t}\,) \bar{P}_\sigma^{\rm{R}} (\bar{t},t^\prime) \nonumber\\
&-\int_{t^\prime}^t \!\! d\bar{t} \, \bar{K}_{\varepsilon_U}^{<} (\bar{t},t) 
\bar{D}^{\rm{R}} (t,\bar{t}\,) \bar{P}_\sigma^{\rm{R}} (\bar{t},t^\prime)~,
\label{Pr}\\
\frac{\partial}{\partial t} \bar{D}^{\rm{R}} (t,t^\prime) = 
&-\sum_\sigma \int_{t^\prime}^t \!\! d\bar{t} \, \bar{K}_{\varepsilon_U}^{>} (t,\bar{t}\,) 
\bar{P}_{-\sigma}^{\rm{R}} (t,\bar{t}\,) \bar{D}^{\rm{R}} (\bar{t},t^\prime)~,
\label{Dr}
\end{align}
and
\begin{widetext}
\begin{align}
\frac{\partial}{\partial t} \bar{E}^{<} (t,t^\prime) 
&= \sum_\sigma \int^{t^\prime}_{-\infty} \!\! d\bar{t} \, \bar{K}_{\varepsilon_0}^{>} (\bar{t},t) 
\bar{P}_\sigma^{<} (t,\bar{t}\,) \left[\bar{E}^{\rm{R}} (t^\prime,\bar{t}\,)\right]^* 
- \sum_\sigma\int^{t}_{-\infty} d\bar{t} \bar{K}_{\varepsilon_0}^{<} (\bar{t},t) 
\bar{P}_\sigma^{\rm{R}} (t,\bar{t}\,) \bar{E}^{<} (\bar{t},t^\prime)~,
\label{Elt}\\
\frac{\partial}{\partial t} \bar{P}_\sigma^{<} (t,t^\prime) 
&= \int^{t^\prime}_{-\infty} \!\! d\bar{t} \, \bar{K}_{\varepsilon_0}^{<} (t,\bar{t}\,) 
\bar{E}^{<}(t,\bar{t}\,) \left[\bar{P}_\sigma^{\rm{R}} (t^\prime,\bar{t}\,)\right]^* 
+ \int^{t^\prime}_{-\infty} \!\! d\bar{t} \, \bar{K}_{\varepsilon_U}^{>} (\bar{t},t) 
\bar{D}^{<} (t,\bar{t}\,) \left[\bar{P}_\sigma^{\rm{R}} (t^\prime,\bar{t}\,)\right]^*
\nonumber\\
&- \int^{t}_{-\infty} \!\! d\bar{t} \, \bar{K}_{\varepsilon_0}^{>} (t,\bar{t}\,) 
\bar{E}^{\rm{R}} (t,\bar{t}\,) \bar{P}_\sigma^{<} (\bar{t},t^\prime) 
- \int^{t}_{-\infty} \!\! d\bar{t} \,  \bar{K}_{\varepsilon_U}^{<} (\bar{t},t) 
\bar{D}^{\rm{R}} (t,\bar{t}\,) \bar{P}_\sigma^{<} (\bar{t},t^\prime)~,
\label{Plt}\\
\frac{\partial}{\partial t} \bar{D}^{<} (t,t^\prime) 
&= \sum_\sigma\int^{t^\prime}_{-\infty} \!\! d\bar{t} \, \bar{K}_{\varepsilon_U}^{<} (t,\bar{t}\,) 
\bar{P}_{-\sigma}^{<} (t,\bar{t}\,) \left[\bar{D}^{\rm{R}} (t^\prime,\bar{t}\,)\right]^* 
- \sum_\sigma\int^{t}_{-\infty} \!\! d\bar{t} \, \bar{K}_{\varepsilon_U}^{>} (t,\bar{t}\,) 
\bar{P}_{-\sigma}^{\rm{R}} (t,\bar{t}\,) \bar{D}^{<} (\bar{t},t^\prime)~
\label{Dlt}
\end{align}
\end{widetext}
with
\begin{align}
\bar{K}_\varepsilon^{\gtrless} (t,t') = \sqrt{\Gamma_{\varepsilon(t)} (t) 
\Gamma_{\varepsilon(t^\prime)} (t^\prime)} \bar{f}_{\varepsilon}^{\,\gtrless} (t,t')
\label{Kfct}
\end{align}
and
\begin{align}
\bar{f}_\varepsilon^{\,\gtrless} (t,t') = \exp\left[i\!\int_{t^\prime}^t \!\! d\bar{t} \, 
\varepsilon(\bar{t}\,)\right] f^{\,\gtrless} (t-t^\prime)~,
\end{align}
where $\varepsilon (t)$ is either $ \varepsilon_0 (t)$ or $ \varepsilon_U (t)$ and 
$f^<(t)=1-f^>(t)$ is the Fourier transform of the Fermi function $f^<(\varepsilon)$ defined by 
\begin{align}
f^< (t) = \int \frac{d\varepsilon}{2\pi} f^< (\varepsilon) \exp\left[ - i \varepsilon t \right]~
\end{align}
with the energy integration taken over the conduction band.
The temperature dependence, which is of main interest, is contained in the integral
kernels $\bar{K}_\varepsilon^{\gtrless} (t,t')$ defined by Eq.~\eqref{Kfct}. In the appendix,
where the details of the derivation of Eqs.~\eqref{Er}--\eqref{Dlt} can be found, we explain 
how these functions enter the formalism.

The initial conditions for Eqs.~\eqref{Er}--\eqref{Dlt} depend on the particular scattering process and how
it is modelled. In our case, the initial conditions are 
\begin{align}
\bar{E}^{\rm{R}} (t,t) = \bar{P}_\sigma^{\rm{R}} (t,t) = \bar{D}^{\rm{R}} (t,t) = 1~,
\label{BCr}
\end{align}
and
\begin{align}
E^< (-\infty,-\infty) &= n_e (-\infty) = 0~,
\label{BCelt}\\
P_\sigma^< (-\infty,-\infty) &= n_{p_\sigma} (-\infty) = \delta_{\sigma,1/2}~,
\label{BCplt}\\
D^< (-\infty,-\infty) &= n_d(-\infty) = 0~.
\label{BCdlt}
\end{align}

Once the equations of motions are solved on a two-dimensional time-grid 
the instantaneous (pseudo) occurrence probabilities for the ${\rm Sr}^{2+}$, $\rm{Sr}^+$,
and $\rm{Sr}^0$ configuration are simply given by the equal-time Green functions,
\begin{align}
n_e (t) &= \bar{E}^< (t,t)~,\label{occ_ne}\\
n_{p_\sigma} (t) &= \bar{P}^<_\sigma (t,t)~, \label{occ_np}\\
n_d (t) &= \bar{D}^< (t,t)~\label{occ_nd}.
\end{align}
Hence, in the notation of pseudo-particles, the 
neutralization probability 
\begin{align}
\alpha = n_d (\infty)~,
\end{align}
that is, it is the probability of double occupancy after completion of the trajectory.

Nordlander and coworkers~\cite{SNL96,SLN94a,SLN94b,LN91} also derived master equations  
for the occurrence probabilities by approximating the time integrals in the Dyson 
equations for the Green functions. Depending on the level of sophistication they 
obtained what they called simple master equations and generalized master equations. 
In the appendix we state the two sets of master equations arising from 
Eqs.~\eqref{Er}--\eqref{Dlt} by adapting this strategy. The reduction of the set of
Dyson equations to a set of master equations utilizes the fact that the functions
$\bar{f}_\varepsilon^\gtrless(t,\bar{t}\,)$ localize the self-energies 
around the time-diagonal. Thus, provided the Green functions vary not too strongly,
they can be put in front of the time integrals. Mathematically, this leads to 
the constraint,~\cite{LN91} 
\begin{align}
R_{0,U}(z)=\Bigg|\frac{v\alpha_{0,U}(z) - 2 \Gamma_{\varepsilon_0,\varepsilon_U}(z)}
{\varepsilon_{0,U}(z)-\varepsilon_F}\Bigg| \ll 1~,
\label{Rv}
\end{align}
where $v$ is the projectile velocity. The functions $\alpha_{0,U}(z)$ are defined by 
requiring $\Gamma_{0,U}(z)=\Delta_{0,U}\exp(-\alpha_{0,U}(z) z)$ which leads to nearly 
constant values for $\alpha_{0,U}$ verifying thereby the exponential dependence of 
our level widths. For the upper level the inequality obviously breaks down at the 
$z=z_c$ where it crosses the Fermi energy. As shown by Langreth and Nordlander~\cite{LN91} 
the master equations can still be used at this point if essentially no charge is 
transferred during the time span the level crosses the Fermi energy. This leads to an 
additional criterion at $z=z_c$. In the next section we will see however that for the upper 
level of the Sr:Au system investigated by He and Yarmoff~\cite{HY10,HY11} the 
constraint~\eqref{Rv} is violated not only at $z=z_c$ but for almost the whole trajectory. 
Hence, in order to analyze the correlation-driven local-moment physics possibly at work in 
this experiment the solutions of the full quantum-kinetic equations are needed.

The physical Green functions $G_\sigma^\lessgtr$ needed for the calculation of the instantaneous 
spectral densities can be constructed from the standard definition of the less-than and 
greater-than Green function~\cite{KB62} by replacing the original electron operators $c_\sigma$ 
and $c^\dagger_\sigma$ by 
pseudo-particle operators according to Eqs.~\eqref{PPRa}--\eqref{PPRb}, neglecting vertex corrections, and 
projecting onto the physical subspace $Q=1$. Thus,   
$G_\sigma^< (t,t^\prime) = \bigl< c_\sigma^\dagger (t^\prime) c_\sigma^\phdag (t) \bigr>$,
for instance, becomes
\begin{align}
G_\sigma^< (t,t^\prime) 
&= \bigl< p_\sigma^\dagger (t^\prime) e (t^\prime) e^\dagger (t) p_\sigma^\phdag (t) \bigr> \nonumber \\
&+ \bigl< d^\dagger (t^\prime) p_{-\sigma}^\phdag (t^\prime) p_{-\sigma}^\dagger (t) d (t) \bigr>~,
\end{align}
which upon employing $E^{\rm{R}} (t,t^\prime) = -i \theta (t-t^\prime) e^{\rm{R}} (t,t^\prime)$ and 
$P_\sigma^{\rm{R}} (t,t^\prime) = -i \theta (t-t^\prime) p_\sigma^{\rm{R}} (t,t^\prime)$ reduces to 
\begin{align}
G_\sigma^< (t,t^\prime) 
&= P_\sigma^< (t,t^\prime) \bigl( e^{\rm{R}} (t^\prime,t) + E^< (t^\prime,t) \bigr) \nonumber\\
&+ D^< (t,t^\prime) \bigl( p_{-\sigma}^{\rm{R}} (t^\prime,t) - P_{-\sigma}^< (t^\prime,t) \bigr)~,
\end{align}
where the products $P_\sigma^< (t,t^\prime) E^< (t^\prime,t)$ and 
$D^< (t,t^\prime) P^<_{-\sigma} (t^\prime,t)$ are of order $Q^2$ and must thus be projected out
to yield 
\begin{align}
G_\sigma^< (t,t^\prime) = P_\sigma^< (t,t^\prime) e^{\rm{R}} (t^\prime,t) 
+ D^< (t,t^\prime) p_{-\sigma}^{\rm{R}} (t^\prime,t)~.
\label{Gless}
\end{align}
A similar calculation leads to 
\begin{align}
G_\sigma^> (t,t^\prime) = p_\sigma^{\rm{R}} (t,t^\prime) E^< (t^\prime,t) 
+ d^{\rm{R}} (t,t^\prime) P_{-\sigma}^< (t^\prime,t)~, 
\label{Ggtr}
\end{align}
where $D^{\rm{R}} (t,t^\prime) = -i \theta (t-t^\prime) d^{\rm{R}} (t,t^\prime)$ has been used.
Note, in the derivation of the formulae for the physical Green functions we introduced Green 
functions $e^{\rm R}$, $p^{\rm R}_\sigma$, and $d^{\rm R}$, which, in contrast to the Green 
functions defined in Eq.~\eqref{factorR} are retarded Green functions with only the Heaviside 
function split-off but the phase factor arising from the on-side energies still included.

The spectral densities for removing or adding at time $T$ a physical electron with 
energy $\omega$ can be obtained from Eq.~(\ref{Gless}) and Eq.~(\ref{Ggtr}) by using
difference variables $T=(t+t^\prime)/2$ and $\tau = t-t^\prime$. A Fourier transformation
with respect to $\tau = t-t^\prime$ yields  
\begin{align}
\rho_\sigma^\lessgtr (\omega, T) = \dfrac{1}{2\pi} \int_{-\infty}^{\infty} \!\! d\tau \, 
G_\sigma^\lessgtr (T+\tau/2, T-\tau/2) \, e^{ i \omega \tau }~.
\end{align}
The normalization of the spectral densities, 
\begin{align}
\int_{-\infty}^\infty \!\! d\omega \, \rho_\sigma^< (\omega, T) &= n_{p_\sigma} (T) + n_d (T) ~,\\
\int_{-\infty}^\infty \!\! d\omega \, \rho_\sigma^> (\omega, T) &= n_{p_{-\sigma}} (T) + n_e (T) ~
\end{align}
is given by the instantaneous occupation of the projectile with a physical electron or a 
physical hole, respectively, written in terms of the occurrence probabilities introduced above. 
This follows directly from the equal-time limit of Eqs.~\eqref{Gless} and~\eqref{Ggtr} by using 
$e^{\rm{R}}(t,t)=d^{\rm R}(t,t)=p_\sigma^{\rm{R}}(t,t)=1$. 
\begin{figure}[t]
\includegraphics[width=0.75\linewidth]{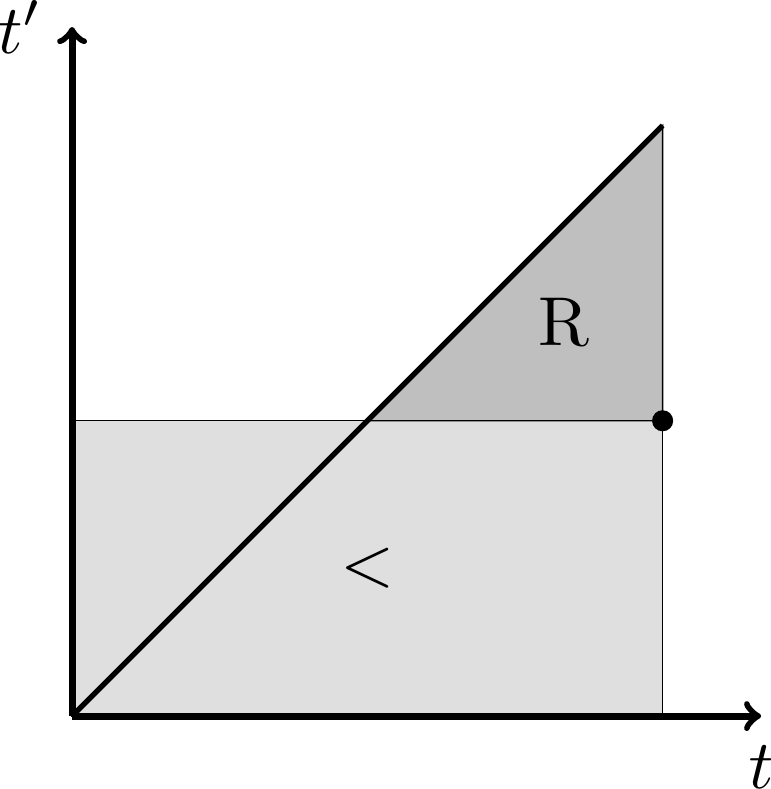}
\caption{Sketch of the numerical scheme used to solve the double-time equations of motion
\eqref{Er}--\eqref{Dlt}. The triangle marks the region in the two-dimensional time-grid
in which the entries of retarded Green functions have to be known in order to calculate
retarded Green functions at the point indicated by the bullet. Likewise, the rectangle 
marks the region in which the entries of (retarded and less-than) Green functions are 
required in order to compute less-than Green functions at the point indicated by the bullet. 
}
\label{numericPic}
\end{figure}

At the end of this section let us say a few words about the numerics required to solve the 
two-dimensional integro-differential equations~\eqref{Er}--\eqref{Dlt}. The discretization strategy 
proposed by Shao and coworkers~\cite{SLN94a} for $U=\infty$ can be also employed for 
finite $U$. The main difference is that two more Green functions have to be calculated on
the time-grid: $\bar{D}^{\rm R}$ and $\bar{D}^<$. The particular structure of the time-integrals 
leads to the numerical strategy shown in Fig.~\ref{numericPic}. First, the retarded Green functions
are calculated, starting from the time-diagonal where their values are simply set to unity
because of the initial condition and then working through the grid points which are on lines 
parallel to the time-diagonal. To compute retarded Green functions at $(t,t^\prime)$ only the 
points in the dark triangle depicted in Fig.~\ref{numericPic} have to be sampled. The calculation 
of the less-than Green functions requires a slightly different scheme. Here the computation 
first proceeds in the $t$- and then in the $t^\prime$-direction, starting from 
$(-t_{\rm max},-t_{\rm max})$ where the initial condition can be employed and redoing this until 
one arrives at the desired grid point. Only grid points in the bright rectangular region of 
Fig.~\ref{numericPic} contribute then to the calculation of less-than functions at the
point $(t,t^\prime)$. 

The computations are time- and memory-consuming. We employ grid-sizes of up to 
$3000\times 3000$. Taking advantage of the symmetry of the Green functions, 
the Green functions in the upper-half of the grid can be obtained from the Green 
functions of the lower-half by complex conjugation which reduces memory space and 
number of calculations by one-half. Even then, however, the calculation of one 
trajectory requires on a $2000\times 2000$ time-grid including the computation of
the level widths eight hours of processing time and $400$ Mb memory on a single
core. To obtain the temperature dependence of the neutralization probability we 
let the projectile run through the trajectory for fifty different temperatures. 
Fortunately, the final charge-state is surprisingly robust against a reduction of 
the size of the time-grid. Empirically we found the neutralization probability (but
not necessarily the occurrence probabilities at intermediate times) to be 
converged already for a $1000\times 1000$ time-grid. A run for a single temperature 
requires then only half an hour making an investigation of the 
temperature dependence of the neutralization process feasible.

\section{Results}
\label{Results}

We now present numerical results. Besides the material parameters listed in 
Table~\ref{Parameters} which should be quite realistic for the Sr:Au 
system investigated by He and Yarmoff we need the turning point $z_{\rm TP}$ 
and the velocity $v$ of the strontium projectile. The radius 
of a strontium atom is around $2.2$ \AA ngstr\"om. It is thus very 
unlikely for the strontium projectile to come closer to the surface than 4-5 
Bohr radii. In atomic units, measuring length in Bohr radii and energy in Hartrees, 
which we use below if not indicated otherwise, we set therefore $z_{\rm TP}=5$. 
For the velocity we take the experimentally determined post-collision velocity
for the whole trajectory, since it is known that due to loss of memory~\cite{OM96}
the outgoing branch determines the final charge-state of the projectile. In atomic 
units, $v=0.134$.~\cite{HY10} 
%
\begin{figure}[t]
\includegraphics[width=0.9\linewidth]{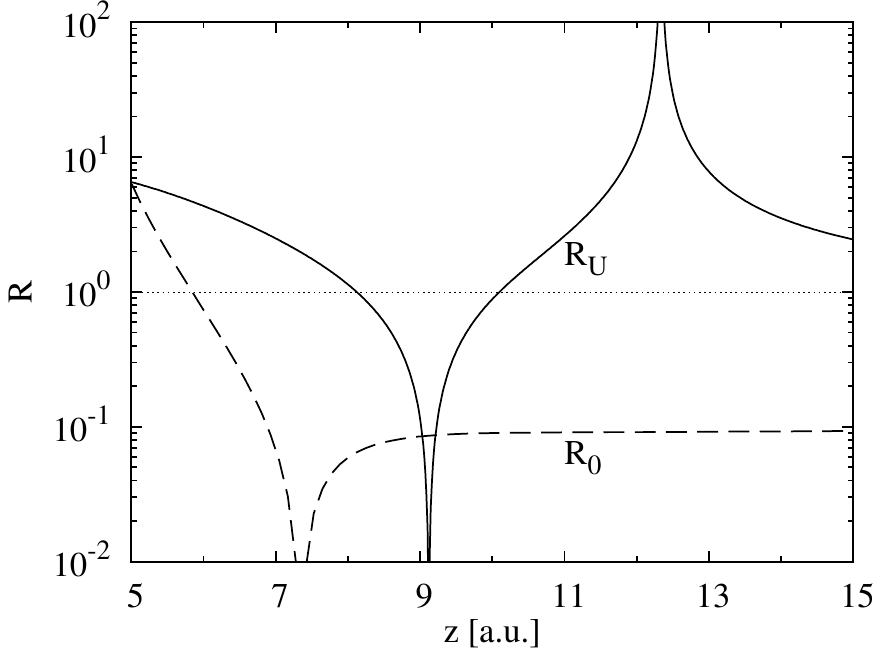}
\caption{The constraints $R_{0}$ and $R_{U}$ as a function of $z$ for the Sr:Au
system investigated by He and Yarmoff.~\cite{HY10,HY11} The approximations
reducing the double-time quantum kinetics of the Dyson equations to either
a set of simple or generalized master equations (see appendix) are valid
only for $R_{0,U}\ll 1$. Hence, for $\varepsilon_0$ master equations could 
be used for $z>6$. But for $\varepsilon_U$ master equations break down
for almost the whole trajectory except for the narrow interval around $z\approx 9$ 
where the vanishing of the numerator in Eq.~\eqref{Rv} leads to small values for $R_{U}$.
The peak in $R_{U}$ around $z\approx 12$ signals the point where
$\varepsilon_U$ crosses the Fermi energy.
}
\label{SCA}
\end{figure}

First, we investigate if the He-Yarmoff experiment~\cite{HY10,HY11} can be described by 
the numerically less demanding master equations (either the simple or the generalized set, 
see appendix). As pointed out in the previous section the master equations should 
provide a reasonable description of the charge transfer if $R_{0,U}(z)\ll 1$. In 
Fig.~\ref{SCA} we plot $R_{0,U}(z)$ for $v=0.0134$ and the level widths and energies 
obtained in Sect.~\ref{Model}. While for the second ionization level $\varepsilon_0$ master 
equations could be in fact used all the way down to $z\approx 6$. For the first ionization 
level $\varepsilon_U$ master equations break down not only at the point where the 
level crosses the Fermi energy but also close to the turning point, where the level width 
turns out to be too large, and far away from the surface, where the projectile velocity is 
too high for the master equations to be applicable. Only in a narrow interval around 
$z\approx 9$, where the high velocity is compensated by the level broadening leading to 
a small numerator in Eq.~\eqref{Rv}, $R_U$ is small 
enough to justify master equations also for $\varepsilon_U$. Since the two ionization 
levels are coupled and the charge transfer occurs not only in the narrow range where 
master equations are applicable to both levels this implies that neither the simple nor the 
generalized master equations can be used to analyze the Sr:Au system investigated by He and 
Yarmoff. Instead, the full double-time quantum-kinetics has to be implemented.

\begin{figure}[t]
\includegraphics[width=0.9\linewidth]{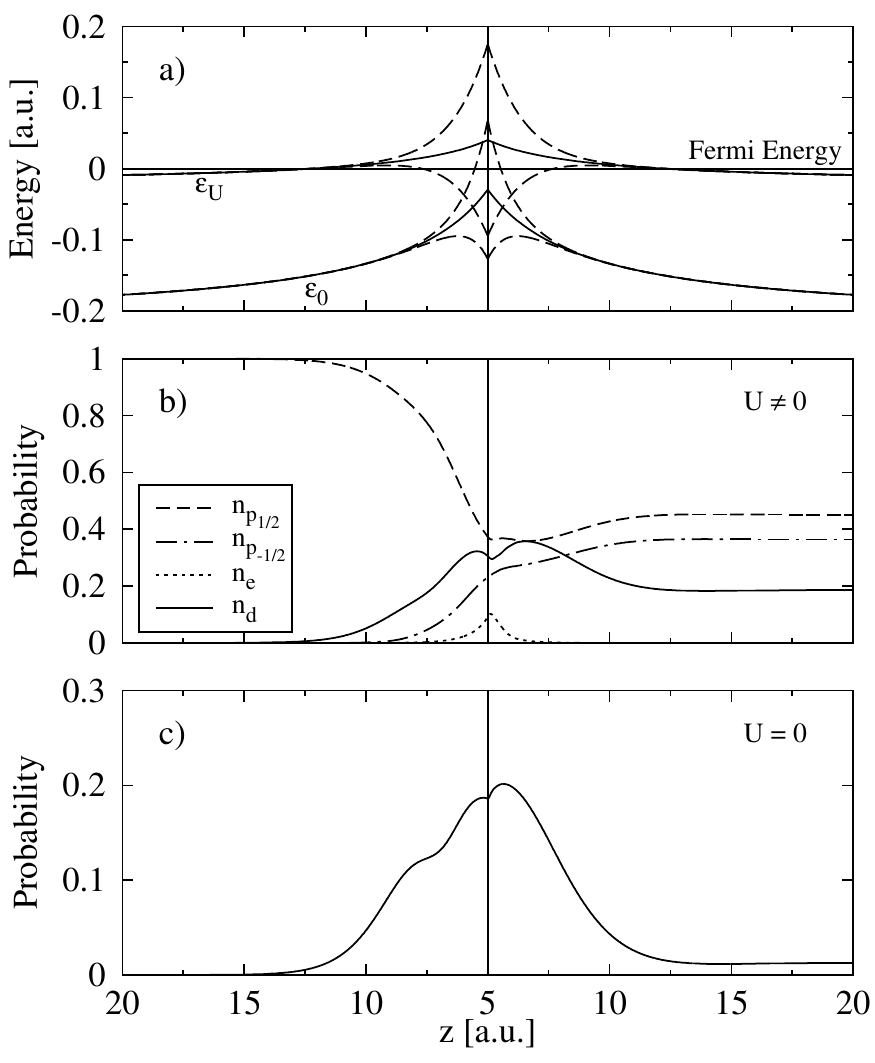}
\caption{Instantaneous physical quantities along the trajectory. The strontium projectile starts
on the left as an ion at a distance $z=20$ with a velocity $v=0.0134$, reaches the turning point
at $z_{\rm TP}=5$ and approaches $z=20$ again thereafter on the right. (a) Energy level diagram. Both
levels (solid lines) are broadened according to $\varepsilon \pm \Gamma$ (dashed lines) with the
instantaneous $\Gamma$ shown in Fig.~\ref{GammaPic}. (b) Occurrence probabilities at $T_s=400$ K
for $\mathrm{Sr^+}$ (dashed and dot-dashed lines), $\mathrm{Sr^0}$ (solid line)
and $\mathrm{Sr^{2+}}$ (dotted line) as obtained from the
finite$-U$ model. The neutralization probability in this case is $\alpha = n_d(20)=0.185$. (c)
Occurrence probability of $\mathrm{Sr^0}$ as obtained from the uncorrelated $U=0$
model which keeps only the first ionization level, that is, the upper on-site energy $\varepsilon_U$.
In this case the $\alpha=0.01$. For other surface temperatures $T_s$ the results look
similar.}
\label{instantan}
\end{figure}

Let us now trace--based on the numerical solution of the double-time Dyson equations--for 
a fixed surface temperature $T_s=400$ K important physical quantities 
while the projectile is on its way through the trajectory. Figure~\ref{instantan} shows in the 
upper panel the shift and broadening of the ionization levels $\varepsilon_U$ and $\varepsilon_0$ 
while the middle panel depicts the instantaneous occurrence probabilities $n_e$, $n_{p_{\pm 1/2}}$, 
and $n_d$ for the $\rm{Sr}^{2+}$, $\rm{Sr}^+$ and $\rm{Sr}^0$ configurations, respectively. The 
projectile starts at $z=20$ on the left, moves along the incoming branch towards the turning point 
$z=5$ from which it returns on the outgoing branch again to the distance $z=20$. The 
strontium projectile starts in the $\rm{Sr}^+$ configuration. Thus only the $\varepsilon_0$ level is 
occupied while the $ \varepsilon_U$ level is empty. During the collision both levels shift upward 
and broaden. The upper level crosses the Fermi energy at $z=z_c\approx 12$.
In the course of the collision the occupation probabilities change and the projectile has a certain 
chance to be at the end in a different charge-state than initially. For the run plotted in 
Fig.~\ref{instantan} the probability for double occupancy at the end, that is, the probability for 
neutralization is $\alpha=n_d(20)=0.185$. For comparison, we show in the lower panel the instantaneous 
occupation of $\varepsilon_U$ as it is obtained when only this level is kept in the modeling, that is, 
for a single-level, uncorrelated $U=0$ model. In this case, the neutralization probability 
$\alpha=0.01$, that is, one order of magnitude smaller.

The physics behind the results shown in Fig.~\ref{instantan} is as follows. Let us first 
focus on the first ionization level. Initially, $\varepsilon_U$, is below the Fermi energy. 
Hence, energetically, not the ionic $\rm{Sr}^+$ but the neutral
$\rm{Sr}^0$ configuration is actually favored. However, as can be seen from the 
vanishing broadening of the level, far away from the surface charge-transfer is negligible. The 
approaching ion is thus initially stabilized due to lack of coupling. When the coupling becomes 
stronger for smaller distances $\varepsilon_U$ crosses however the Fermi energy. The ion is then
energetically stabilized. Roughly speaking, the first ionization level has a chance to capture an 
electron from the metal only when $|E_F-\varepsilon_U(t)|<\Gamma_U (t)$; in the notation of Sosolik 
and coworkers the Sr:Au system is in the coupling-dominated regime.~\cite{SHL03} From the upper 
panel in Fig.~\ref{instantan} we see that this is the case only for a very small portion of the 
trajectory, close to the turning point. As a result, the neutralization probability 
$\alpha$ should be in any case much smaller than unity as indeed it is. 
Due to the thermal broadening of the target's Fermi edge the efficiency of electron-capture 
into the first ionization level increases with temperature. Thus, if this was the only process 
involved in the charge-transfer, the neutralization probability should monotonously increase 
with temperature, contrary to the experimental data which initially increase and then 
decrease (see below). The charge-transfer must be thus more involved. 
Indeed, as can be seen in the upper panel in Fig.~\ref{instantan}, the second ionization 
level $\varepsilon_0$ comes also close to the Fermi energy. In those parts of the trajectory 
where $|E_F-\varepsilon_0 (t)|<\Gamma_0 (t)$ it is thus conceivable that the electron initially 
occupying $\varepsilon_0$ may leave the projectile. That is, holes may transfer from the 
surface to the second ionization level thereby compensating the electron-transfer into 
the first. The hole-transfer, absent in the uncorrelated $U=0$ model, tendentiously 
favors the ion with increasing temperature and should by itself lead to a neutralization 
probability decreasing with temperature. 

That during the collision the ionization levels of strontium come so close to the Fermi 
energy of the gold target, with the first one crossing it and the second one coming so close 
to it to enable hole-transfer, led He and Yarmoff to suggest that the neutralization process
is driven by electron correlations. The experimentally found negative temperature dependence
of $\alpha$ above $T_s = 600$ K strengthened their conclusion. It agrees qualitatively with 
what Merino and Marston predicted theoretically on the basis of a correlated-electron 
model for the neutralization of calcium ions on copper surfaces.~\cite{MM98} The work of 
Shao and coworkers~\cite{SNL96} suggested moreover that the negative temperature 
dependence of $\alpha$ is caused by a mixed-valence resonance transiently formed in the course 
of the collision.
\begin{figure}[t]
\includegraphics[width=0.9\linewidth]{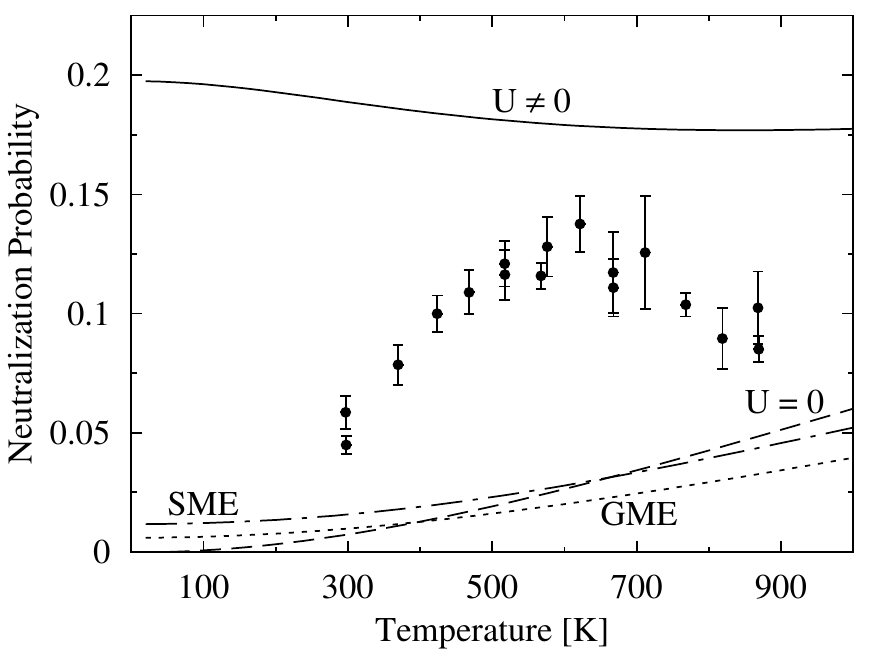}
\caption{Temperature dependence of the neutralization probability
$\alpha=n_d(20)$ for a $\mathrm{Sr^+}$ ion hitting with $v=0.0134$
a gold surface. The turning point $z_{\rm TP}=5$. Also shown are
the data of He and Yarmoff.~\cite{HY10} The solid and long-dashed
lines are for the finite$-U$ and the uncorrelated $U=0$ model, respectively,
showing that correlations enhance the neutralization probability to the 
experimental order of magnitude. By moving the turning point farther away 
from the surface we could make the results for $U\neq 0$ to overlap
with the experimental data. However, we do not use $z_{\rm TP}$ as a fit parameter 
for reasons explained in the main text. The non-monotonous temperature 
dependence of the experimental data cannot be reproduced regardless of the 
value of the turning point. The dashed-dotted and the dotted line are
the neutralization probabilities arising, respectively, from the numerical 
solution of the set of simple or the set of generalized master equations
given in the appendix.
}
\label{temperature}
\end{figure}

After these qualitative remarks we now discuss the temperature dependence of the 
neutralization probability quantitatively. In Fig.~\ref{temperature} we show 
the experimental data of He and Yarmoff~\cite{HY10} and compare it with our theoretical 
results. For the parameters of Table~\ref{Parameters} the theoretical 
neutralization probability (solid line) turns out a bit too large but it is still of 
the correct order of magnitude indicating that the material parameters as well as the 
procedures for calculating the level widths are reasonable. In contrast to the
experimental data we find however over the whole temperature range only a weak negative 
temperature dependence. Also plotted in Fig.~\ref{temperature} is the temperature dependence 
of the neutralization probability arising from the uncorrelated $U=0$ model (long-dashed 
line) and--for completeness--the one obtained from the numerical solution of either the set 
of simple (dashed-dotted line) or the set of generalized master equations (dotted lines)
listed at the end of the appendix. 

Clearly, without correlations the neutralization probability is too small indicating that correlations 
play an important role in the charge-transfer from the gold target to the strontium projectile. 
The chosen turning point $z_{\rm TP}=5$ is in fact most favorable for the uncorrelated model. 
In reality the turning point may be farther away from the surface. A larger value of $z_{\rm TP}$ 
leads however to smaller neutralization probabilities. Hence, the results for the uncorrelated 
model would be pushed farther away from the experimental data while the results for the correlated 
model would come closer to it. We hesitate however to use $z_{\rm TP}$ as a fit parameter 
because of the shortcomings of the finite$-U$ non-crossing approximation discussed in the 
next section. 

The neutralization probabilities arising from the master equations are also much smaller than 
the ones obtained from the full quantum kinetics. Decreasing the turning point would push 
them of course closer to the experimental data (without reproducing the non-monotonous 
temperature dependence). However, the numerical values for $R_{0,U}(z)$ shown in 
Fig.~\ref{SCA} indicate that the approximations leading to the master equations 
cannot be justified. Hence, the results for $\alpha$ obtained from the master equations 
should not be artificially pushed towards experimental data by manipulating the turning 
point. Instead one should--if at all--try to push the correlated $U\neq 0$ data closer to 
the experimental data by changing the parameters of the Sr:Au system within physically 
sensible bounds. 
\begin{figure*}[t]
\begin{minipage}{0.5\linewidth}
\includegraphics[width=0.9\linewidth]{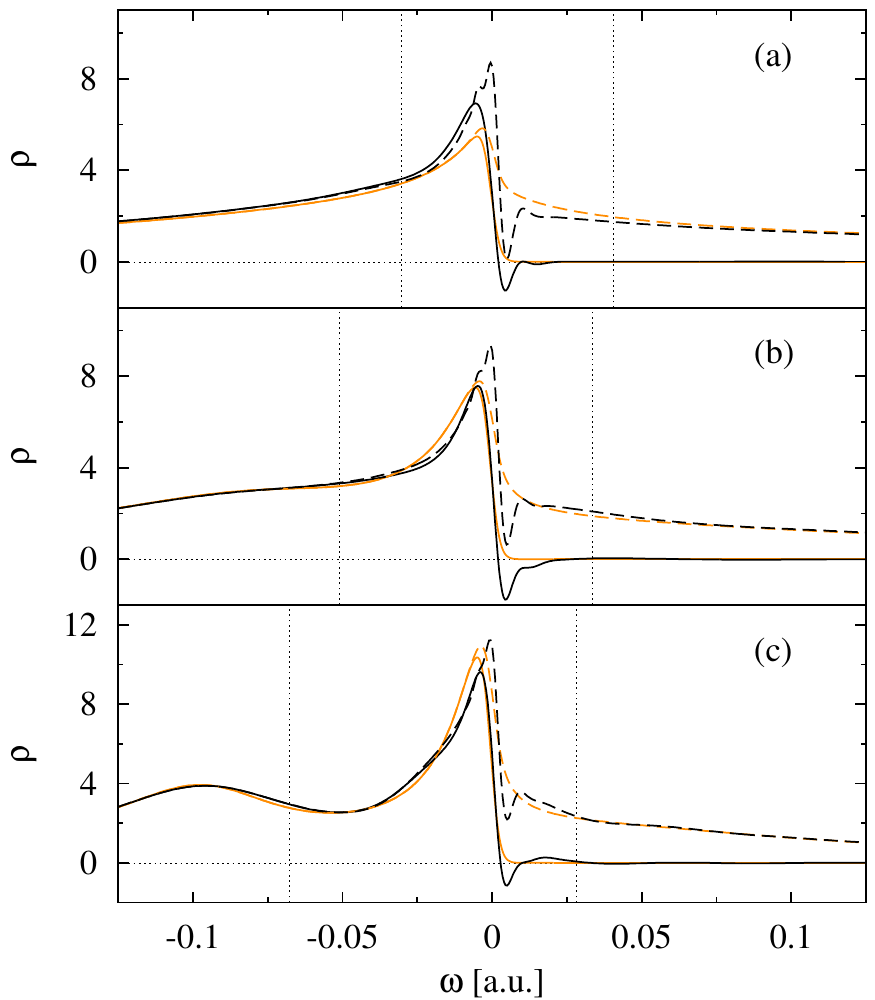}
\end{minipage}\begin{minipage}{0.5\linewidth}
\includegraphics[width=0.9\linewidth]{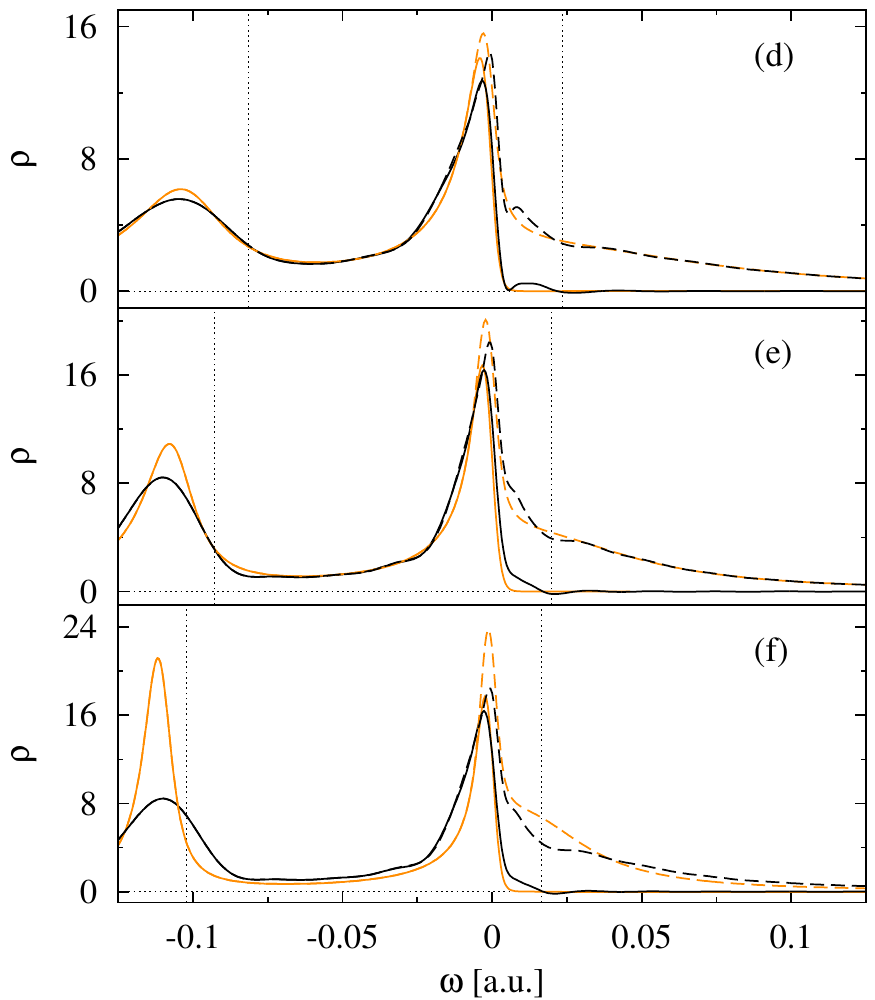}
\end{minipage}
\caption{(Color online) Projectile spectral densities summed over the two spin
orientations, respectively, at $z=5.0$, $5.5$, $6.0$, $6.5$,
$7.0$, and $7.5$ on the outgoing branch of the trajectory [panel (a) to (f)] for a
$\mathrm{Sr^+}$ ion hitting with $v=0.0134$ a gold surface at temperature $T_s=400$ K.
For other surface temperatures in the range relevant for the experiment the spectral functions look qualitatively 
similar. The black lines correspond to the instantaneous spectral densities at the given positions
with the solid lines denoting the occupied and the dashed lines the total spectral densities.
The vertical dotted lines indicate the instantaneous positions of the on-site energies $\varepsilon_0$
and $\varepsilon_U$ while the orange lines give the equilibrated occupied (solid lines) and total
(dashed lines) spectral densities at the corresponding positions. The target's Fermi energy
is located at $\omega=0$.}
\label{SpectralFct}
\end{figure*}

Any attempt however to improve the theoretical data by changing the material parameters 
and hence the single-particle matrix elements of the Anderson-Newns Hamiltonian was unsuccessful.
A slight increase of the metal's work function from $\phi=5.1$~eV to $\phi=5.15$~eV, 
for instance, decreased the neutralization rate but eliminated at the 
same time the weak negative temperature dependence. Decreasing the work function from 
$\phi=5.1$~eV to $\phi=5.05$~eV, on the other hand, increased the 
theoretical neutralization rate but did also not lead to a stronger negative temperature 
dependence let alone to a non-monotonous one. Changing the turning 
point $z_{\rm TP}$ affects the neutralization probability as indicated in the previous 
paragraph but again wipes out the weak negative temperature dependence. The effect of 
the Doppler broadening~\cite{PLN99,Winter02,SHL03} we did not investigate. We take 
all this as an indication that the correlation effects encoded in the finite-$U$ 
non-crossing approximation are too fragile. Going beyond this approximation is thus 
unavoidable. 

Another observation should be mentioned. The starting point $z=20$ can be relatively 
freely chosen. If it is closer to the surface the slopes of the instantaneous occurrence 
probabilities in Fig.~\ref{instantan} are steeper so that there is hardly 
any difference in the probabilities at the turning point and no difference at the end of 
the trajectory. As a result, the final neutralization probability is independent of 
the precise starting conditions. The loss of memory in charge-transferring atom-surface
collisions has been also found by Onufriev and Marston.~\cite{OM96} It justifies 
using the pre-collision velocity for the whole trajectory.

In the region where charge-transfer is strongest the two ionization levels overlap. The 
absence of energy separation together with the conditional temporal weighting due to the 
dynamics of the collision process makes it very hard to tell a priori whether electron- 
or hole-transfer dominates the outcome of the collision. Simply changing the matrix 
elements of the Anderson-Newns model in the hope to reproduce the experimentally found 
temperature anomaly is pointless as we have indeed seen. Even more so, since the hypothesized 
electron correlations of the local-moment type strongly distort the projectile's density of 
states in the vicinity of the target's Fermi energy. Any attempt to guess the projectile's 
final charge-state on the basis of the single-particle quantities shown in the upper
panel of Fig.~\ref{instantan} has thus to fail. In order to see whether 
the weak negative temperature dependence of $\alpha$ is already a qualitative hint
for a mixed-valence scenario to be at work in the neutralization of strontium ions on 
gold surfaces we calculated therefore the instantaneous spectral densities for the 
projectile. If local-moment physics is present these functions should feature transient 
resonances at the target's Fermi energy. 

In Fig.~\ref{SpectralFct} we present for a selected set of distances along the outgoing 
branch of the trajectory and for $T_s=400$~K the instantaneous spectral densities summed 
over the two spin orientations. The occupied part of the spectral densities (solid black lines),
that is, the spectrally resolved probability for removing a physical electron, as well as 
the total spectral densities (dashed black lines), which in addition contain also the 
spectrally resolved probability for adding an electron, are shown. For orientation we 
also plot the equilibrated spectral densities (solid and dashed orange lines) which we 
obtained by fixing the widths and energetic positions of the levels 
to the values at that particular distance and then letting the system evolve in time 
up to the point where it reaches a quasi-stationary state. The negative values of the
instantaneous spectral densities close to and at the turning point should not be interpreted 
too literally. First, we cannot rule out that in the numerical Fourier transformation the Gibbs 
phenomenon occurs although the results for the equilibrated spectral densities speak 
against it. Second, and most importantly, the instantaneous spectral densities 
$\rho_\sigma^\lessgtr (\omega, T)$ are Wigner distributions in 
energy $\omega$ and time $T$. These two quantities, however, cannot be measured 
simultaneously. Usually Wigner distributions deal with quantum-mechanical uncertainties by 
becoming negative in some regions of the space in which they are defined.~\cite{Baker58}
Integrated over energy, that is, the zeroth-order moments of the Wigner distributions 
$\rho_\sigma^\lessgtr (\omega, T)$ give however always the correct occupancies at the particular
time as can be easily checked by a comparison with the data obtained from the integration 
of the equations of motion.

Let us start with panel (a) of Fig.~\ref{SpectralFct} which shows the spectral densities at 
the closest encounter. The overlapping ionization levels are very broad at this point leading 
however to a spectral density which due to electronic correlations is enhanced at the 
Fermi energy $\omega=0$. The uncorrelated model would not give this enhancement. Moving 
outwards (panels (b)--(f)) the spectral densities change, because of the decreasing widths 
and the shifting of the ionization levels, developing in addition to the resonance at $\omega=0$
features in the vicinity of the two instantaneous ionization levels which are indicated
by the two vertical dotted lines. Although the additional structure due to the upper 
ionization level is only a high-energy shoulder to the peak at $\omega=0$ the spectral 
densities develop the shape expected from a quantum impurity: Two charge-transfer peaks 
and a resonance at the Fermi energy. This can be most clearly seen in panel (f). Since 
close to the surface the upper charge-transfer peak merges more or less with the peak 
at the Fermi energy to form a mixed-valence resonance the Sr:Au system is in the 
mixed-valence regime.

The dominating spectral feature at all the distances shown in Fig.~\ref{SpectralFct} is the 
enhancement at the target's Fermi energy. Despite the quantitative discrepancies between 
the measured and the computed neutralization probabilities our theoretical results for the 
spectral densities suggest--for realistic single-particle parameters and without
any fit parameter--that local-moment physics is present in the 
Sr:Au system and may thus control the neutralization of ${\rm Sr}^+$ on Au surfaces 
as anticipated by He and Yarmoff.~\cite{HY10} More specifically, from Fig.~\ref{instantan} 
we read-off that most of the charge-transfer occurs between the turning point $z_{\rm TP}=5$ 
and the crossing point $z_c\approx 12$ whereas from Fig.~\ref{SpectralFct} we see that for 
these distances the Sr:Au system develops at the Fermi energy of the Au target a 
mixed-valence resonance with a high-energy tail 
varying on the scale of the thermal energy. The weak negative temperature dependence we obtain 
for $\alpha$ is thus due to the mixed-valance resonance in the projectile's spectral density in 
accordance with what Merino and Marston predicted for the correlated Ca:Cu system.~\cite{MM98} 
The comparison in Fig.~\ref{temperature} with the results obtained from the uncorrelated $U=0$ 
model suggests moreover that it is also the mixed-valence resonance which enhances the 
neutralization probabilities to the experimentally found order of magnitude. 

Obviously, our results support He and Yarmoff's mixed-valence scenario~\cite{HY10,HY11} 
only qualitatively but not quantitatively. Either the transient local-moment correlations 
are too weak, occur at the wrong distance, or are simply too short-lived. It requires further 
theoretical work to tell which one of these possibilities applies. 

\section{Conclusions}
\label{Conclusions}

We presented a realistically parameterized Anderson-Newns model for charge-transferring collisions 
between a strontium projectile and a gold target and used the model to analyze from a many-body 
theoretical point of view the experiment of He and Yarmoff~\cite{HY11,HY10} which indicated 
that in this type of surface collision a mixed-valence resonance affects the final 
charge-state of the projectile.

In contrast to the measured neutralization probability which 
initially increases and then decreases with temperature the computed data show only 
the correlation-induced enhancement, making the calculated neutralization probability 
of the correct order of magnitude, and a weak negative temperature dependence.
The analysis of the projectile's instantaneous spectral densities revealed however
that both the enhancement and the negative temperature dependence arise from a 
mixed-valence resonance at the target's Fermi energy in qualitative agreement with 
what Merino and Marston found for the Ca:Cu system,~\cite{MM98} which is another 
projectile-target combination which could display local-moment physics. Thus, qualitatively, 
our results support He and Yarmoff's interpretation of their data in terms of a 
mixed-valence resonance.

We followed the theoretical approach of Nordlander and coworkers.~\cite{SNL96,SLN94a,SLN94b,LN91} 
It is based on the non-crossing approximation
for Anderson-impurity-type models and contour-ordered Green functions.
That we do not find the anomalous temperature dependence of the neutralization probability 
while having a transient mixed-valence resonance in the instantaneous spectral densities could 
have two reasons. 
First, the accuracy of the semi-empirical estimates we developed 
for the single-particle matrix elements of the Anderson-Newns Hamiltonian may be not enough.
The shift of the two ionization levels was obtained from classical considerations based 
on image charges while the width of the levels was computed from Hartree-Fock and step-potential 
wave functions. Ab-initio calculations or measurements of these two quantities would be very
helpful, in particular, for distances close to the turning point. Second, the finite$-U$ 
non-crossing approximation most probably does not yield the correct energy scale 
of the resonance transiently formed at the Fermi energy of the target. Indeed, for finite $U$
the non-crossing approximation does not self-consistently sum-up all leading terms in $1/N$ 
where $N=2$ is the degeneracy of the 5s level. In equilibrium it is known that the non-crossing 
approximation underestimates due to this inconsistency the width of the Kondo resonance 
considerable.~\cite{Sakai88} Systematically summing-up all diagrams to leading order by the 
one-crossing approximation~\cite{OK06,KW05,PG89,HS89,Sakai88} remedies this shortcoming as 
does the dynamical $1/N$ approximation used by Merino and Marston~\cite{MM98} and equation
of motion approaches working directly with the physical Green functions defining the 
spectral densities.~\cite{GFM05} It should be
also noted that the temperature anomaly occurs over an interval of only $600$~K corresponding
to an energy interval $\Delta E\approx 0.002$ in atomic units. The spectral features in the 
vicinity of the Fermi energy which drive the anomaly have thus to be known with an energy 
resolution better than $10^{-3}$. 

Specifically our results for the spectral densities make us adhere to the 
mixed-valence scenario. 
Besides the above mentioned improvements on the theoretical side 
further experimental analysis would be however also required to clarify the issue. The 
velocity dependence of the effect, for instance, would be of great interest because it is the 
projectile velocity which determines whether the instantaneous correlations get frozen-in 
and manifest themselves in the final charge-state of the projectile.
We would thus expect the experimentally observed temperature anomaly 
to depend strongly on the projectile's velocity. Changing the work function
and the collision geometry would be also of interest. The former manipulates the point 
where the upper ionization level crosses the target's Fermi energy whereas the latter 
changes the effective temperature via Doppler broadening.
The temperature anomaly of the neutralization probability should hence also depend on the 
work function of the surface and the angle of incident. 

It may be easier to realize local-moment physics in electrically biased 
semiconductor nanostructures
but demonstrating it to be also present in charge-transferring atom-surface collisions 
may open up avenues for further research which are not yet anticipated. The Sr:Au system
investigated by He and Yarmoff may well be a very promising candidate.

\section*{Acknowledgements}
M. P. was funded by the federal state of Mecklenburg-Western Pomerania through a 
postgraduate scholarship within the International Helmholtz Graduate School for Plasma Physics. In addition, 
support from the Deutsche Forschungsgemeinschaft through project B10 of the Transregional Collaborative 
Research Center SFB/TRR24 is greatly acknowledged.\\

\section*{Appendix}
In this appendix we lay out the basic definitions and notations we used in setting up the quantum
kinetic equations \eqref{Er}--\eqref{Dlt} of Sec.~\ref{QuantumKinetics}. The equations have been
originally derived by Shao and coworkers~\cite{SLN94b}. As in our previous work on the de-excitation 
of meta-stable molecules at surfaces,~\cite{MBF12b} we stay as closely as possible to the notation 
of Nordlander and coworkers~\cite{SLN94a,SLN94b,LN91} and deviate from it only when it improves the
readability of the equations. 

Contour-ordered Green functions~\cite{KB62,Keldysh65} describing the empty, the single occupied, 
and the double occupied projectile, 
\begin{align}
i E(t,t^\prime) & = \bigl\langle T_\mathcal{C} \, e(t) \, e^\dag(t^\prime) \bigr\rangle \;, \\
i P_\sigma(t,t^\prime) & = \bigl\langle T_\mathcal{C} \, p_\sigma^\phdag(t) \, p_\sigma^\dag(t^\prime) \bigr\rangle \;, \\
i D(t,t^\prime) & = \bigl\langle T_\mathcal{C} \, d(t) \, d^\dag(t^\prime) \bigr\rangle ~, 
\end{align}
as well as metal electrons, 
\begin{align}
i G_{\vec{k},\sigma}(t,t^\prime)  = 
\bigl\langle T_\mathcal{C} \, c_{\vec{k},\sigma}^\phdag(t) \, 
c_{\vec{k},\sigma}^\dag(t^\prime) 
\bigr\rangle~,
\end{align}
where the brackets denote the statistical average with respect to the initial density matrix,
constitute the basis of the formalism. The functions $D$ and $E$ are bosonic propagators while $P_\sigma$
and $G_{\vec{k},\sigma}$ are fermionic. For any of the four Green functions listed above the analytic 
pieces, that is, the less-than and the greater-than functions, are given by 
\begin{align}
i H(t,t^\prime) & = \Theta_\mathcal{C}(t-t^\prime) \, H^>(t,t^\prime) 
                  \mp \Theta_\mathcal{C}(t^\prime-t) \, H^<(t,t^\prime) \;,
\end{align}
where $H$ stands for $E$, $P_\sigma$, $D$, or $G_{\vec{k},\sigma}$ and $\Theta_\mathcal{C}$ is the Heaviside 
function defined on the complex time contour. The upper sign holds for fermionic and the lower sign for bosonic 
Green functions. As usual, the corresponding retarded functions read
\begin{align}
i H^R(t,t^\prime) = \theta(t-t^\prime)\bigl[ H^>(t,t^\prime) \pm H^<(t,t^\prime)\bigr]~,
\end{align}
where again the upper (lower) sign holds for fermionic (bosonic) functions and $\theta$ is now the 
Heaviside function on the real time axis. 

Similarly, the self-energies $\Sigma_\sigma$, $\Pi_{\rm e}$, and $\Pi_{\rm d}$ for the single occupied, 
the empty, and the double occupied projectile can be split into analytic pieces which in turn give rise 
to retarded self-energies, 
\begin{align}
i \Sigma_\sigma^{\rm R}(t,t^\prime) = \theta(t-t^\prime)\bigl[ \Sigma_\sigma^>(t,t^\prime) 
+ \Sigma_\sigma^<(t,t^\prime)\bigr]~,\\
i \Pi_{\rm e,d}^{\rm R}(t,t^\prime) = \theta(t-t^\prime)\bigl[ \Pi_{\rm e,d}^>(t,t^\prime) 
- \Pi_{\rm e,d}^<(t,t^\prime)\bigr]~.
\end{align}
Within the non-crossing approximation the metal electrons are undressed. Hence, below
$G_{\vec{k},\sigma}$ is always the bare propagator and no self-energy has to be specified 
for the metal electrons.~\cite{SNL96,SLN94a,SLN94b,LN91}

On the real-time axis the analytic pieces of the Green function obey the set of Dyson equations
($\hbar=1$): 
\begin{widetext}
\begin{align}
i\frac{\partial}{\partial t}E^{\rm{R}} (t,t^\prime) &=\delta(t-t^\prime)
+\int_{-\infty}^\infty \!\! d\bar{t} \, \Pi^{\rm R}_{\rm e}(t,\bar{t}\,)
E^{\rm{R}}(\bar{t},t^\prime)~
\label{Dyson_Er}\\
\Bigl[i\frac{\partial}{\partial t}-\varepsilon_0(t)\Bigl] P^{\rm{R}}_\sigma (t,t^\prime) &=\delta(t-t^\prime)
+\int_{-\infty}^\infty \!\! d\bar{t} \, \Sigma_\sigma^{\rm R}(t,\bar{t}\,)
P_\sigma^{\rm{R}}(\bar{t},t^\prime)~
\\
\Bigl[i\frac{\partial}{\partial t}-\varepsilon_0(t)-\varepsilon_U(t)\Bigl] D^{\rm{R}} (t,t^\prime) &=\delta(t-t^\prime)
+\int_{-\infty}^\infty \!\! d\bar{t} \, \Pi_{\rm d}^{\rm R}(\bar{t},t)
D^{\rm{R}}(\bar{t},t^\prime)
\end{align}
\begin{align}
i\frac{\partial}{\partial t} E^{<} (t,t^\prime)
&= \int^{\infty}_{-\infty} \!\! d\bar{t} \, \Pi_{\rm e}^{\rm R}(t,\bar{t}\,)
E^{<} (\bar{t},t^\prime)
+\int^{\infty}_{-\infty} d\bar{t}  \, \Pi_{\rm e}^{<}(t,\bar{t}\,)
E^{\rm A} (\bar{t},t^\prime)
\\
\Bigl[i\frac{\partial}{\partial t}-\varepsilon_0(t)\Bigl] P_\sigma^{<} (t,t^\prime)
&= \int^{\infty}_{-\infty} \!\! d\bar{t} \, \Sigma_\sigma^{\rm R}(t,\bar{t}\,)
P_\sigma^{<} (\bar{t},t^\prime)
+\int^{\infty}_{-\infty} d\bar{t}  \, \Sigma_\sigma^{<}(t,\bar{t}\,)
P_\sigma^{\rm A} (\bar{t},t^\prime)
\\
\Bigl[i\frac{\partial}{\partial t}-\varepsilon_0(t)-\varepsilon_U(t)\Bigl] D^{<} (t,t^\prime)
&= \int^{\infty}_{-\infty} \!\! d\bar{t} \, \Pi_{\rm d}^{\rm R}(t,\bar{t}\,)
D^{<} (\bar{t},t^\prime)
+\int^{\infty}_{-\infty} d\bar{t}  \, \Pi_{\rm d}^{<}(t,\bar{t}\,)
D^{\rm A} (\bar{t},t^\prime)~.
\label{Dyson_Dlt}
\end{align}
\end{widetext}

The self-energies in the non-crossing approximation are shown in Fig.~\ref{selfPic}, where 
the self-energy $\Sigma_\sigma$ for the single occupied projectile is split into two pieces, 
$\Sigma_{\sigma,0}$ and $\Sigma_{\sigma,U}$, depending on whether the empty or the double 
occupied state appears as a virtual state. Applying standard diagrammatic rules~\cite{LL81} together 
with the Langreth-Wilkins rules~\cite{LW72} given in our notation in Ref.~\cite{MBF12b} yields after 
projection to the $Q=1$ subspace~\cite{LN91,WM94,AL03} the following mathematical expressions for 
the analytic pieces of the self-energies:
\begin{align}
\Pi^\gtrless_{\rm d}(t,t^\prime) &= \sum_\sigma \int \frac{d\varepsilon}{2\pi}  
                              K_{\varepsilon}^\gtrless(t,t^\prime) P_{-\sigma}^\gtrless(t,t^\prime)~,
\label{Pilt}\\
\Pi^{\rm R}_{\rm d}(t,t^\prime)        &= \sum_\sigma \int \frac{d\varepsilon}{2\pi}
                              K_{\varepsilon}^>(t,t^\prime) P_{-\sigma}^{\rm R}(t,t^\prime)~,\\
\Pi^\gtrless_{\rm e}(t,t^\prime) &= \sum_\sigma \int \frac{d\varepsilon}{2\pi} 
                              K_{\varepsilon}^\lessgtr(t^\prime,t) P_\sigma^\gtrless(t,t^\prime)~,\\
\Pi^{\rm R}_{\rm e}(t,t^\prime)        &= \sum_\sigma \int \frac{d\varepsilon}{2\pi}
                              K_{\varepsilon}^<(t^\prime,t) P_\sigma^{\rm R}(t,t^\prime)~,
\end{align}
\begin{align}
\Sigma^\gtrless_{\sigma,0}(t,t^\prime) &= \int \frac{d\varepsilon}{2\pi} 
                              K_{\varepsilon}^\gtrless(t,t^\prime) E^\gtrless(t,t^\prime)~,\\
\Sigma^{\rm R}_{\sigma,0}(t,t^\prime) &= \int \frac{d\varepsilon}{2\pi}
                              K_{\varepsilon}^>(t,t^\prime) E^{\rm R}(t,t^\prime)~,\\
\Sigma^\gtrless_{\sigma,U}(t,t^\prime) &= \int \frac{d\varepsilon}{2\pi} 
                              K_{\varepsilon}^\lessgtr(t^\prime,t) D^\gtrless(t,t^\prime)~\\
\Sigma^{\rm R}_{\sigma,U}(t,t^\prime) &=\int \frac{d\varepsilon}{2\pi} 
                              K_{\varepsilon}^<(t^\prime,t) D^{\rm R}(t,t^\prime)~
\label{SigmaR}
\end{align}
with 
\begin{align}
K_\varepsilon^\gtrless(t,t^\prime)=\sqrt{\Gamma_\varepsilon(t)\Gamma_\varepsilon(t^\prime)}
f^\gtrless(\varepsilon)e^{-i\varepsilon(t-t^\prime)}~,
\label{Kunbarred}
\end{align}
where $\varepsilon$ is an energy variable to be integrated over.

In obtaining the self-energies we took advantage of the fact that the propagator of the
metal electrons is undressed and spin independent. As a result, $\Sigma_{\sigma,0}$ 
and $\Sigma_{\sigma,U}$ (and thus $\Sigma_{\sigma}$) are independent of the electron 
spin. Furthermore, we assumed the tunneling matrix element $V_{\vec{k}}(t)$ to 
factorize in the variables $t$ and $\vec{k}$. In our case this is approximately true 
since the strongest time dependence in Eq.~\eqref{Vkfinal} comes from the modified 
Bessel function $K_\alpha$ giving rise to a nearly exponential time dependence of 
$V_{\vec{k}}(t)$. The function
\begin{align}
\Gamma_\varepsilon(t,t^\prime) &= 2\pi \sum_{\vec{k}} V_{\vec{k}}(t) V^*_{\vec{k}}(t^\prime)
\delta(\varepsilon-\varepsilon_{\vec{k}})
\end{align}
initially appearing in the self-energies can thus be approximately rewritten as~\cite{SLN94a,LN91}
\begin{align}
\Gamma_\varepsilon(t,t^\prime)\simeq \sqrt{\Gamma_\varepsilon(t)\Gamma_\varepsilon(t^\prime)}
\end{align}
with $\Gamma_\varepsilon(t)$ defined by Eq.~\eqref{GammaEq} leading eventually to the expressions 
for the self-energies given above.

Inserting the self-energies \eqref{Pilt}--\eqref{SigmaR} into the Dyson equations
\eqref{Dyson_Er}--\eqref{Dyson_Dlt} and rewriting the equations in terms of the reduced 
Green functions defined by Eqs.~\eqref{factorR} and \eqref{factorLess} yields after an
approximate $\varepsilon$-integration Eqs. \eqref{Er}--\eqref{Dlt} of 
Sec.~\ref{QuantumKinetics}. 

Due to the approximate $\varepsilon$-integration the functions 
$\bar{K}_\varepsilon^{\gtrless}(t,t')$ enter the formalism. In the definition~\eqref{Kfct} 
of these functions the subscript $\varepsilon$ denotes not an energy variable but the 
functional dependence on $\varepsilon(t)$. To see this consider the Dyson equation~\eqref{Dyson_Er}. 
In terms of reduced Green functions it reads
\begin{widetext}
\begin{align}
\partial_t \bar{E}^{\rm R}(t,t^\prime) &= -\sum_\sigma \int_{t^\prime}^t d\bar{t} 
              \int \frac{d\varepsilon}{2\pi} 
              \sqrt{\Gamma_\varepsilon(t)\Gamma_\varepsilon(\bar{t}\,)}f^<(\varepsilon)
              \exp\left[-i\!\int_{\bar{t}}^t \!\! d\tau \,
              (\varepsilon_0(\tau)-\varepsilon)\right] 
              \bar{P}_{\sigma}^{\rm R}(t,\bar{t}\,)\bar{E}^{\rm R}(\bar{t},t^\prime) \label{Eaux}\\
              &\simeq -\sum_\sigma \int_{t^\prime}^t d\bar{t}
              \sqrt{\Gamma_{\varepsilon_0(t)}(t)\Gamma_{\varepsilon_0(\bar{t}\,)}(\bar{t}\,)}
              f^<(\bar{t}-t) \exp\left[-i\!\int_{\bar{t}}^t \!\! d\tau \,
              \varepsilon_0(\tau)\right] 
              \bar{P}_{\sigma}^{\rm R}(t,\bar{t}\,)\bar{E}^{\rm R}(\bar{t},t^\prime) \label{intuitive}\\
              &= -\sum_\sigma \int_{t^\prime}^t d\bar{t}
              \sqrt{\Gamma_{\varepsilon_0(t)}(t)\Gamma_{\varepsilon_0(\bar{t}\,)}(\bar{t}\,)}
              \bar{f}_{\varepsilon_0}^<(\bar{t},t) 
              \bar{P}_{\sigma}^{\rm R}(t,\bar{t}\,)\bar{E}^{\rm R}(\bar{t},t^\prime) \\
              &= -\sum_\sigma \int_{t^\prime}^t d\bar{t} 
              \bar{K}_{\varepsilon_0}^<(\bar{t},t) 
              \bar{P}_{\sigma}^{\rm R}(t,\bar{t}\,)\bar{E}^{\rm R}(\bar{t},t^\prime)
\end{align}

\end{widetext}

with $\bar{K}_{\varepsilon_0}^<(\bar{t},t)$ as defined in Eq.~\eqref{Kfct}. The step 
from the first to the second line involves the approximate $\varepsilon$-integration
resulting in the Fourier transformation of the Fermi function and in fixing 
the energy variables of the level widths as indicated. We did not attempt to derive
it mathematically by an asymptotic stationary-phase analysis.~\cite{BH86} Instead we followed 
Shao and coworkers~\cite{SLN94b} and adopted a qualitative, physics-based reasoning. It yields 
the very intuitive equation~\eqref{intuitive} and reduces moreover the numerical 
effort considerably because it is no longer necessary to perform at each time-grit point 
$(t,t^\prime)$ an $\varepsilon$-integration. Alternatively $\Gamma_\varepsilon(t)$ could 
be replaced in~\eqref{Eaux} by an average over the energy range of the conduction 
band and then put in front of the $\varepsilon$-integral.~\cite{SLN94a} But this seems 
to be even more ad-hoc.

Similar manipulations can be performed for the other Dyson equations. At the end one 
obtains equations~\eqref{Er}--\eqref{Dlt} of Sec.~\ref{QuantumKinetics}. The equations
are identical to the ones given by Shao and coworkers in the book 
edited by Rabalais~\cite{SLN94b} if--as we did--the pseudo-particle
operator $p_\sigma$ is taken to be fermionic.

The kinetic equations~\eqref{Er}--\eqref{Dlt} are a complicated set of two-dimensional 
integro-differential equations. Nordlander and coworkers~\cite{SLN94a,SLN94b,LN91} 
showed however that in situations where the functions $\bar{f}_\varepsilon^\gtrless(t,\bar{t}\,)$ 
and hence the self-energies are sufficiently peaked at $t=\bar{t}$ the Dyson equations
for the less-than Green functions can be reduced to master equations for the occurrence 
probabilities which are numerically less expensive. Depending on whether 
retarded Green functions are taken at equal times and hence pushed in front
of the time integrals or not two sets of master equations can be derived: the simple and 
the generalized master equations.~\cite{SLN94a,SLN94b,LN91} Applying this reasoning to 
Eqs.~\eqref{Er}--\eqref{Dlt} yields at the level where retarded Green functions are taken at 
equal times a set of simple master equations, 
\begin{widetext}
\begin{align}
\dfrac{d}{dt} n_e(t) = &-2 \Gamma_0(t) f^<\bigl(\varepsilon_0(t)\bigr) n_e(t) 
+ \Gamma_0(t) f^>\bigl(\varepsilon_0(t)\bigr) \Bigl( n_{p_{1/2}}(t) + n_{p_{-1/2}}(t) \Bigr)~, \\
\dfrac{d}{dt} n_{p_\sigma}(t) = &- \Bigl( \Gamma_0 (t) f^>\bigl(\varepsilon_0(t)\bigr) + \Gamma_U (t) 
f^<\bigl(\varepsilon_U(t)\bigr) \Bigr) n_{p_\sigma} (t) \nonumber\\&+ \Gamma_0(t) f^<\bigl(\varepsilon_0(t)\bigr) n_e(t) 
+ \Gamma_U(t) f^>\bigl(\varepsilon_U(t)\bigr) n_d(t)~, \\
\dfrac{d}{dt} n_d(t) = &-2 \Gamma_U(t) f^>\bigl(\varepsilon_U(t)\bigr) n_d(t) 
+ \Gamma_U(t) f^<\bigl(\varepsilon_U(t)\bigr) \Bigl( n_{p_{1/2}}(t) + n_{p_{-1/2}}(t) \Bigr)~,
\end{align}
and at the advanced level, where retarded Green functions are kept non-diagonal in time, 
a set of generalized master equations, 
\begin{align}
\dfrac{d}{dt} n_e(t) = &-2 n_e(t) \sum_\sigma \int_{-\infty}^t \!\! d\bar{t} \, 
\Im\Bigl(  \bar{K}_{\varepsilon_0}^{<} (\bar{t}, t) \bar{P}_\sigma^{\rm{R}} (t,\bar{t}\,) 
\Bigr) + 2 \sum_\sigma n_{p_\sigma}(t) \int_{-\infty}^t\!\! d\bar{t} \, 
\Im\Bigl( \bar{K}_{\varepsilon_0}^{>} (\bar{t}, t) \left[\bar{E}^{\rm{R}} (t,\bar{t}\,)\right]^* \Bigr)~,\\
\dfrac{d}{dt} n_{p_\sigma} (t) = &-2 n_{p_\sigma}(t) \int_{-\infty}^t \!\! d\bar{t} \, 
\Im\Bigl( \bar{K}_{\varepsilon_0}^{>} (t, \bar{t}\,) \bar{E}^{\rm{R}} (t,\bar{t}\,) + 
\bar{K}_{\varepsilon_U}^{<} (\bar{t}, t) \bar{D}^{\rm{R}} (t,\bar{t}\,) \Bigr) \nonumber\\
&+ 2 n_e(t) \int_{-\infty}^t \!\! d\bar{t} \, \Im\Bigl( \bar{K}_{\varepsilon_0}^{<} (t, \bar{t}\,) 
\left[\bar{P}_\sigma^{\rm{R}} (t,\bar{t}\,)\right]^* \Bigr) + 2 n_d(t) \int_{-\infty}^t \!\! 
d\bar{t} \, \Im\Bigl( \bar{K}_{\varepsilon_U}^{>} (\bar{t}, t) 
\left[\bar{P}_\sigma^{\rm{R}} (t,\bar{t}\,)\right]^* \Bigr)~,\\
\dfrac{d}{dt} n_d(t) = &-2 n_d(t) \sum_\sigma \int_{-\infty}^t \!\! d\bar{t} \, 
\Im\Bigl(  \bar{K}_{\varepsilon_U}^{>} (t, \bar{t}\,) \bar{P}_\sigma^{\rm{R}} (t,\bar{t}\,) \Bigr) 
+ 2 \sum_\sigma n_{p_\sigma}(t) \int_{-\infty}^t\!\! d\bar{t} \, \Im\Bigl( \bar{K}_{\varepsilon_U}^{<} (t, \bar{t}\,) 
\left[\bar{D}^{\rm{R}} (t,\bar{t}\,)\right]^* \Bigr)~
\end{align}
with occurrence probabilities $n_e(t)$, $n_{p_\sigma}(t)$, and $n_d(t)$ as defined in               
Eqs.~\eqref{occ_ne}--\eqref{occ_nd}. The retarded Green functions required in the generalized 
master equations can be obtained by utilizing the localization of 
$\bar{f}_\varepsilon^\gtrless(t,\bar{t}\,)$ around the time-diagonal also in the Dyson equations for 
the retarded Green functions. As a result one obtains, 
\begin{align}
\bar{E}^{\rm{R}} (t,t^\prime) &= \exp\left[-\sum_\sigma \int_{t^\prime}^t \!\! d\tau \int_{t^\prime}^{\tau} \!\! d\bar{t} \, \bar{K}_{\varepsilon_0}^{<} (\bar{t},\tau) \bar{P}_\sigma^{\rm{R}} (\tau,\bar{t}\,) \right]~,\\
\bar{P}^{\rm{R}}_\sigma (t,t^\prime) &= \exp\left[-\int_{t^\prime}^t \!\! d\tau \int_{t^\prime}^{\tau} \!\! d\bar{t} \, \Bigl(\bar{K}_{\varepsilon_0}^{>} (\tau,\bar{t}\,) \bar{E}^{\rm{R}} (\tau,\bar{t}\,) + \bar{K}_{\varepsilon_U}^{<} (\bar{t}, \tau) \bar{D}^{\rm{R}} (\tau,\bar{t}\,) \Bigr) \right]~,\\
\bar{D}^{\rm{R}} (t,t^\prime) &= \exp\left[-\sum_\sigma \int_{t^\prime}^t \!\! d\tau \int_{t^\prime}^{\tau}
\!\! d\bar{t} \, \bar{K}_{\varepsilon_U}^{>} (\tau, \bar{t}\,) \bar{P}_{-\sigma}^{\rm{R}} (\tau,\bar{t}\,) \right]~.
\end{align}

\end{widetext}

A rigorous determination of the range of validity of these equations by asymptotic 
techniques~\cite{BH86} is complicated because the functions $\bar{f}_\varepsilon^\gtrless(t,\bar{t}\,)$ 
are not only localized around the time-diagonal but also strongly oscillating. Simple saddle-point 
arguments are thus not sufficient but have to be augmented by a stationary-phase analysis. 
Analyzing moreover the whole set of Dyson equations by these techniques seems to be impractical. 
Langreth and Nordlander~\cite{LN91} investigated therefore the validity of the approximations 
empirically and developed qualitative criteria which have to be satisfied for master equations 
to provide a reasonable description of the charge-transfer between the projectile and the 
target surface. As shown in Sect.~\ref{Results} the basic constraint~\eqref{Rv} they developed 
is not satisfied for the Sr:Au system investigated by He and Yarmoff.~\cite{HY10,HY11} The full 
double-time quantum kinetic equations have thus to be solved to analyze this experiment.\\


\begin{thebibliography}{71}
\expandafter\ifx\csname natexlab\endcsname\relax\def\natexlab#1{#1}\fi
\expandafter\ifx\csname bibnamefont\endcsname\relax
  \def\bibnamefont#1{#1}\fi
\expandafter\ifx\csname bibfnamefont\endcsname\relax
  \def\bibfnamefont#1{#1}\fi
\expandafter\ifx\csname citenamefont\endcsname\relax
  \def\citenamefont#1{#1}\fi
\expandafter\ifx\csname url\endcsname\relax
  \def\url#1{\texttt{#1}}\fi
\expandafter\ifx\csname urlprefix\endcsname\relax\def\urlprefix{URL }\fi
\providecommand{\bibinfo}[2]{#2}
\providecommand{\eprint}[2][]{\url{#2}}

\bibitem[{\citenamefont{Monreal}(2014)}]{Monreal14}
\bibinfo{author}{\bibfnamefont{R.~C.} \bibnamefont{Monreal}},
  \bibinfo{journal}{Progr. Surf. Sci.} \textbf{\bibinfo{volume}{89}},
  \bibinfo{pages}{80} (\bibinfo{year}{2014}).

\bibitem[{\citenamefont{Winter and Burgd\"orfer}(2007)}]{Winter07}
\bibinfo{editor}{\bibfnamefont{H.-P.} \bibnamefont{Winter}} \bibnamefont{and}
  \bibinfo{editor}{\bibfnamefont{J.}~\bibnamefont{Burgd\"orfer}}, eds.,
  \emph{\bibinfo{title}{{Slow heavy-particle induced electron emission from
  solid surface}}} (\bibinfo{publisher}{Springer-Verlag},
  \bibinfo{address}{Berlin Heidelberg}, \bibinfo{year}{2007}).

\bibitem[{\citenamefont{Winter}(2002)}]{Winter02}
\bibinfo{author}{\bibfnamefont{H.}~\bibnamefont{Winter}},
  \bibinfo{journal}{Phys. Rep.} \textbf{\bibinfo{volume}{367}},
  \bibinfo{pages}{387} (\bibinfo{year}{2002}).

\bibitem[{\citenamefont{Rabalais}(1994)}]{Rabalais94}
\bibinfo{editor}{\bibfnamefont{J.~W.} \bibnamefont{Rabalais}}, ed.,
  \emph{\bibinfo{title}{{Low energy ion-surface interaction}}}
  (\bibinfo{publisher}{Wiley and Sons}, \bibinfo{address}{New York},
  \bibinfo{year}{1994}).

\bibitem[{\citenamefont{Los and Geerlings}(1990)}]{LG90}
\bibinfo{author}{\bibfnamefont{J.}~\bibnamefont{Los}} \bibnamefont{and}
  \bibinfo{author}{\bibfnamefont{J.~J.~C.} \bibnamefont{Geerlings}},
  \bibinfo{journal}{Phys. Rep.} \textbf{\bibinfo{volume}{190}},
  \bibinfo{pages}{133} (\bibinfo{year}{1990}).

\bibitem[{\citenamefont{Brako and Newns}(1989)}]{BN89}
\bibinfo{author}{\bibfnamefont{R.}~\bibnamefont{Brako}} \bibnamefont{and}
  \bibinfo{author}{\bibfnamefont{D.~M.} \bibnamefont{Newns}},
  \bibinfo{journal}{Rep. Prog. Phys.} \textbf{\bibinfo{volume}{52}},
  \bibinfo{pages}{655} (\bibinfo{year}{1989}).

\bibitem[{\citenamefont{Modinos}(1987)}]{Modinos87}
\bibinfo{author}{\bibfnamefont{A.}~\bibnamefont{Modinos}},
  \bibinfo{journal}{Progr. Surf. Sci.} \textbf{\bibinfo{volume}{26}},
  \bibinfo{pages}{19} (\bibinfo{year}{1987}).

\bibitem[{\citenamefont{Yoshimori and Makoshi}(1986)}]{YM86}
\bibinfo{author}{\bibfnamefont{A.}~\bibnamefont{Yoshimori}} \bibnamefont{and}
  \bibinfo{author}{\bibfnamefont{K.}~\bibnamefont{Makoshi}},
  \bibinfo{journal}{Prog. Surf. Sci.} \textbf{\bibinfo{volume}{21}},
  \bibinfo{pages}{251} (\bibinfo{year}{1986}).

\bibitem[{\citenamefont{Newns et~al.}(1983)\citenamefont{Newns, Makoshi, Brako,
  and van Wunnik}}]{NMB83}
\bibinfo{author}{\bibfnamefont{D.~M.} \bibnamefont{Newns}},
  \bibinfo{author}{\bibfnamefont{K.}~\bibnamefont{Makoshi}},
  \bibinfo{author}{\bibfnamefont{R.}~\bibnamefont{Brako}}, \bibnamefont{and}
  \bibinfo{author}{\bibfnamefont{J.~N.~M.} \bibnamefont{van Wunnik}},
  \bibinfo{journal}{Physica Scripta} \textbf{\bibinfo{volume}{T6}},
  \bibinfo{pages}{5} (\bibinfo{year}{1983}).

\bibitem[{\citenamefont{Kraus et~al.}(2007)\citenamefont{Kraus, Falter, Fantz,
  Franzen, Heinemann, McNeely, Riedl, and SPeth}}]{KFF07}
\bibinfo{author}{\bibfnamefont{W.}~\bibnamefont{Kraus}},
  \bibinfo{author}{\bibfnamefont{H.-D.} \bibnamefont{Falter}},
  \bibinfo{author}{\bibfnamefont{U.}~\bibnamefont{Fantz}},
  \bibinfo{author}{\bibfnamefont{P.}~\bibnamefont{Franzen}},
  \bibinfo{author}{\bibfnamefont{B.}~\bibnamefont{Heinemann}},
  \bibinfo{author}{\bibfnamefont{P.}~\bibnamefont{McNeely}},
  \bibinfo{author}{\bibfnamefont{R.}~\bibnamefont{Riedl}}, \bibnamefont{and}
  \bibinfo{author}{\bibfnamefont{E.}~\bibnamefont{SPeth}},
  \bibinfo{journal}{Rev. Sci. Instrum.} \textbf{\bibinfo{volume}{79}},
  \bibinfo{pages}{02C108} (\bibinfo{year}{2007}).

\bibitem[{\citenamefont{Lieberman and Lichtenberg}(2005)}]{LL05}
\bibinfo{author}{\bibfnamefont{M.~A.} \bibnamefont{Lieberman}}
  \bibnamefont{and} \bibinfo{author}{\bibfnamefont{A.~J.}
  \bibnamefont{Lichtenberg}}, \emph{\bibinfo{title}{{Principles of plasma
  discharges and materials processing}}}
  (\bibinfo{publisher}{Wiley-Interscience}, \bibinfo{address}{New York},
  \bibinfo{year}{2005}).

\bibitem[{\citenamefont{Rabalais}(2003)}]{Rabalais03}
\bibinfo{author}{\bibfnamefont{J.~W.} \bibnamefont{Rabalais}},
  \emph{\bibinfo{title}{Principles and applications of ion scattering
  spectrometry: Surface chemical and structural analysis}}
  (\bibinfo{publisher}{Wiley and Sons}, \bibinfo{address}{New York},
  \bibinfo{year}{2003}).

\bibitem[{\citenamefont{Harada et~al.}(1997)\citenamefont{Harada, Masuda, and
  Ozaki}}]{HMO97}
\bibinfo{author}{\bibfnamefont{Y.}~\bibnamefont{Harada}},
  \bibinfo{author}{\bibfnamefont{S.}~\bibnamefont{Masuda}}, \bibnamefont{and}
  \bibinfo{author}{\bibfnamefont{H.}~\bibnamefont{Ozaki}},
  \bibinfo{journal}{Chem. Rev.} \textbf{\bibinfo{volume}{97}},
  \bibinfo{pages}{1897} (\bibinfo{year}{1997}).

\bibitem[{\citenamefont{Hewson}(1993)}]{Hewson93}
\bibinfo{author}{\bibfnamefont{A.~C.} \bibnamefont{Hewson}},
  \emph{\bibinfo{title}{{The Kondo problem to heavy fermions}}}
  (\bibinfo{publisher}{Cambridge University Press},
  \bibinfo{address}{Cambridge}, \bibinfo{year}{1993}).

\bibitem[{\citenamefont{Fulde}(1995)}]{Fulde95}
\bibinfo{author}{\bibfnamefont{P.}~\bibnamefont{Fulde}},
  \emph{\bibinfo{title}{{Electron correlations in molecules and solids}}}
  (\bibinfo{publisher}{Springer Verlag}, \bibinfo{address}{Berlin},
  \bibinfo{year}{1995}).

\bibitem[{\citenamefont{Chang and Chen}(2009)}]{CC09}
\bibinfo{author}{\bibfnamefont{A.~M.} \bibnamefont{Chang}} \bibnamefont{and}
  \bibinfo{author}{\bibfnamefont{J.~C.} \bibnamefont{Chen}},
  \bibinfo{journal}{Rep. Prog. Phys.} \textbf{\bibinfo{volume}{72}},
  \bibinfo{pages}{096501} (\bibinfo{year}{2009}).

\bibitem[{\citenamefont{Pustilnik and Glazman}(2004)}]{PG04}
\bibinfo{author}{\bibfnamefont{M.}~\bibnamefont{Pustilnik}} \bibnamefont{and}
  \bibinfo{author}{\bibfnamefont{L.}~\bibnamefont{Glazman}},
  \bibinfo{journal}{J. Phys. Condens. Matter} \textbf{\bibinfo{volume}{16}},
  \bibinfo{pages}{R513} (\bibinfo{year}{2004}).

\bibitem[{\citenamefont{Aguado and Langreth}(2003)}]{AL03}
\bibinfo{author}{\bibfnamefont{R.}~\bibnamefont{Aguado}} \bibnamefont{and}
  \bibinfo{author}{\bibfnamefont{D.~C.} \bibnamefont{Langreth}},
  \bibinfo{journal}{Phys. Rev. B} \textbf{\bibinfo{volume}{67}},
  \bibinfo{pages}{245307} (\bibinfo{year}{2003}).

\bibitem[{\citenamefont{Goldhaber-Gordon
  et~al.}(1998{\natexlab{a}})\citenamefont{Goldhaber-Gordon, Shtrikman, Mahalu,
  Abusch-Magder, Meirav, and Kastner}}]{GSM98}
\bibinfo{author}{\bibfnamefont{D.}~\bibnamefont{Goldhaber-Gordon}},
  \bibinfo{author}{\bibfnamefont{H.}~\bibnamefont{Shtrikman}},
  \bibinfo{author}{\bibfnamefont{D.}~\bibnamefont{Mahalu}},
  \bibinfo{author}{\bibfnamefont{D.}~\bibnamefont{Abusch-Magder}},
  \bibinfo{author}{\bibfnamefont{U.}~\bibnamefont{Meirav}}, \bibnamefont{and}
  \bibinfo{author}{\bibfnamefont{M.~A.} \bibnamefont{Kastner}},
  \bibinfo{journal}{Nature} \textbf{\bibinfo{volume}{391}},
  \bibinfo{pages}{156} (\bibinfo{year}{1998}{\natexlab{a}}).

\bibitem[{\citenamefont{Cronenwett et~al.}(1998)\citenamefont{Cronenwett,
  Oosterkamp, and Kouwenhoven}}]{COK98}
\bibinfo{author}{\bibfnamefont{S.~M.} \bibnamefont{Cronenwett}},
  \bibinfo{author}{\bibfnamefont{T.~H.} \bibnamefont{Oosterkamp}},
  \bibnamefont{and} \bibinfo{author}{\bibfnamefont{L.~P.}
  \bibnamefont{Kouwenhoven}}, \bibinfo{journal}{Science}
  \textbf{\bibinfo{volume}{281}}, \bibinfo{pages}{540} (\bibinfo{year}{1998}).

\bibitem[{\citenamefont{Goldhaber-Gordon
  et~al.}(1998{\natexlab{b}})\citenamefont{Goldhaber-Gordon, Gores,
  Kastner, Shtrikman, Mahalu, and Meirav}}]{GGK98}
\bibinfo{author}{\bibfnamefont{D.}~\bibnamefont{Goldhaber-Gordon}},
  \bibinfo{author}{\bibfnamefont{J.}~\bibnamefont{Gores}},
  \bibinfo{author}{\bibfnamefont{M.~A.} \bibnamefont{Kastner}},
  \bibinfo{author}{\bibfnamefont{H.}~\bibnamefont{Shtrikman}},
  \bibinfo{author}{\bibfnamefont{D.}~\bibnamefont{Mahalu}}, \bibnamefont{and}
  \bibinfo{author}{\bibfnamefont{U.}~\bibnamefont{Meirav}},
  \bibinfo{journal}{Phys. Rev. Lett.} \textbf{\bibinfo{volume}{81}},
  \bibinfo{pages}{5225} (\bibinfo{year}{1998}{\natexlab{b}}).

\bibitem[{\citenamefont{Wingreen and Meir}(1994)}]{WM94}
\bibinfo{author}{\bibfnamefont{N.~S.} \bibnamefont{Wingreen}} \bibnamefont{and}
  \bibinfo{author}{\bibfnamefont{Y.}~\bibnamefont{Meir}},
  \bibinfo{journal}{Phys. Rev. B} \textbf{\bibinfo{volume}{49}},
  \bibinfo{pages}{11040} (\bibinfo{year}{1994}).

\bibitem[{\citenamefont{Grabert and Devoret}(1992)}]{GD92}
\bibinfo{editor}{\bibfnamefont{H.}~\bibnamefont{Grabert}} \bibnamefont{and}
  \bibinfo{editor}{\bibfnamefont{M.~H.} \bibnamefont{Devoret}}, eds.,
  \emph{\bibinfo{title}{{Single charge tunneling: Coulomb blockade phenomena in
  nanostructures}}} (\bibinfo{publisher}{Plenum Press}, \bibinfo{address}{New
  York}, \bibinfo{year}{1992}).

\bibitem[{\citenamefont{Nghiem and Costi}(2014)}]{NC14}
\bibinfo{author}{\bibfnamefont{H.~T.~M.} \bibnamefont{Nghiem}}
  \bibnamefont{and} \bibinfo{author}{\bibfnamefont{T.~A.} \bibnamefont{Costi}},
  \bibinfo{journal}{Phys. Rev. B} \textbf{\bibinfo{volume}{90}},
  \bibinfo{pages}{035129} (\bibinfo{year}{2014}).

\bibitem[{\citenamefont{Lechtenberg and Anders}(2014)}]{LA14}
\bibinfo{author}{\bibfnamefont{B.}~\bibnamefont{Lechtenberg}} \bibnamefont{and}
  \bibinfo{author}{\bibfnamefont{F.~B.} \bibnamefont{Anders}},
  \bibinfo{journal}{Phys. Rev. B} \textbf{\bibinfo{volume}{90}},
  \bibinfo{pages}{045117} (\bibinfo{year}{2014}).

\bibitem[{\citenamefont{M\"{u}hlbacher
  et~al.}(2011)\citenamefont{M\"{u}hlbacher, Urban, and Komnik}}]{MUK11}
\bibinfo{author}{\bibfnamefont{L.}~\bibnamefont{M\"{u}hlbacher}},
  \bibinfo{author}{\bibfnamefont{D.~F.} \bibnamefont{Urban}}, \bibnamefont{and}
  \bibinfo{author}{\bibfnamefont{A.}~\bibnamefont{Komnik}},
  \bibinfo{journal}{Phys. Rev. B} \textbf{\bibinfo{volume}{83}},
  \bibinfo{pages}{075107} (\bibinfo{year}{2011}).

\bibitem[{\citenamefont{Cohen and Rabani}(2011)}]{CR11}
\bibinfo{author}{\bibfnamefont{G.}~\bibnamefont{Cohen}} \bibnamefont{and}
  \bibinfo{author}{\bibfnamefont{E.}~\bibnamefont{Rabani}},
  \bibinfo{journal}{Phys. Rev. B} \textbf{\bibinfo{volume}{84}},
  \bibinfo{pages}{075150} (\bibinfo{year}{2011}).

\bibitem[{\citenamefont{Shao et~al.}(1996)\citenamefont{Shao, Nordlander,
  Langreth}}]{SNL96}
\bibinfo{author}{\bibfnamefont{H.}~\bibnamefont{Shao}},
  \bibinfo{author}{\bibfnamefont{P.} \bibnamefont{Nordlander}},
  \bibnamefont{and}
  \bibinfo{author}{\bibfnamefont{D.~C.}~\bibnamefont{Langreth}},
  \bibinfo{journal}{Phys. Rev. Lett.} \textbf{\bibinfo{volume}{77}},
  \bibinfo{pages}{948} (\bibinfo{year}{1996}).

\bibitem[{\citenamefont{Merino and Marston}(1998)}]{MM98}
\bibinfo{author}{\bibfnamefont{J.}~\bibnamefont{Merino}} \bibnamefont{and}
  \bibinfo{author}{\bibfnamefont{J.~B.} \bibnamefont{Marston}},
  \bibinfo{journal}{Phys. Rev. B} \textbf{\bibinfo{volume}{58}},
  \bibinfo{pages}{6982} (\bibinfo{year}{1998}).

\bibitem[{\citenamefont{He and Yarmoff}(2011)}]{HY11}
\bibinfo{author}{\bibfnamefont{X.}~\bibnamefont{He}} \bibnamefont{and}
  \bibinfo{author}{\bibfnamefont{J.~A.} \bibnamefont{Yarmoff}},
  \bibinfo{journal}{Nucl. Instrum. Meth. Phys. Res. B}
  \textbf{\bibinfo{volume}{269}}, \bibinfo{pages}{1195} (\bibinfo{year}{2011}).

\bibitem[{\citenamefont{He and Yarmoff}(2010)}]{HY10}
\bibinfo{author}{\bibfnamefont{X.}~\bibnamefont{He}} \bibnamefont{and}
  \bibinfo{author}{\bibfnamefont{J.~A.} \bibnamefont{Yarmoff}},
  \bibinfo{journal}{Phys. Rev. Lett.} \textbf{\bibinfo{volume}{105}},
  \bibinfo{pages}{176806} (\bibinfo{year}{2010}).

\bibitem[{\citenamefont{Shao et~al.}(1994{\natexlab{a}})\citenamefont{Shao,
  Langreth, and Nordlander}}]{SLN94a}
\bibinfo{author}{\bibfnamefont{H.}~\bibnamefont{Shao}},
  \bibinfo{author}{\bibfnamefont{D.~C.} \bibnamefont{Langreth}},
  \bibnamefont{and}
  \bibinfo{author}{\bibfnamefont{P.}~\bibnamefont{Nordlander}},
  \bibinfo{journal}{Phys. Rev. B} \textbf{\bibinfo{volume}{49}},
  \bibinfo{pages}{13929} (\bibinfo{year}{1994}{\natexlab{a}}).

\bibitem[{\citenamefont{Shao et~al.}(1994{\natexlab{b}})\citenamefont{Shao,
  Langreth, and Nordlander}}]{SLN94b}
\bibinfo{author}{\bibfnamefont{H.}~\bibnamefont{Shao}},
  \bibinfo{author}{\bibfnamefont{D.~C.} \bibnamefont{Langreth}},
  \bibnamefont{and}
  \bibinfo{author}{\bibfnamefont{P.}~\bibnamefont{Nordlander}}, in
  \emph{\bibinfo{booktitle}{{Low energy ion-surface interaction}}}, edited by
  \bibinfo{editor}{\bibfnamefont{J.~W.} \bibnamefont{Rabalais}}
  (\bibinfo{publisher}{Wiley and Sons}, \bibinfo{address}{New York},
  \bibinfo{year}{1994}{\natexlab{b}}), p. \bibinfo{pages}{117}.

\bibitem[{\citenamefont{Langreth and Nordlander}(1991)}]{LN91}
\bibinfo{author}{\bibfnamefont{D.~C.} \bibnamefont{Langreth}} \bibnamefont{and}
  \bibinfo{author}{\bibfnamefont{P.}~\bibnamefont{Nordlander}},
  \bibinfo{journal}{Phys. Rev. B} \textbf{\bibinfo{volume}{43}},
  \bibinfo{pages}{2541} (\bibinfo{year}{1991}).

\bibitem[{\citenamefont{Kasai and Okiji}(1987)}]{KO87}
\bibinfo{author}{\bibfnamefont{H.}~\bibnamefont{Kasai}} \bibnamefont{and}
  \bibinfo{author}{\bibfnamefont{A.}~\bibnamefont{Okiji}},
  \bibinfo{journal}{Surface science} \textbf{\bibinfo{volume}{183}},
  \bibinfo{pages}{147} (\bibinfo{year}{1987}).

\bibitem[{\citenamefont{Nakanishi et~al.}(1988)\citenamefont{Nakanishi, Kasai,
  and Okiji}}]{NKO88}
\bibinfo{author}{\bibfnamefont{H.}~\bibnamefont{Nakanishi}},
  \bibinfo{author}{\bibfnamefont{H.}~\bibnamefont{Kasai}}, \bibnamefont{and}
  \bibinfo{author}{\bibfnamefont{A.}~\bibnamefont{Okiji}},
  \bibinfo{journal}{Surface science} \textbf{\bibinfo{volume}{197}},
  \bibinfo{pages}{515} (\bibinfo{year}{1988}).

\bibitem[{\citenamefont{Romero et~al.}(2009)\citenamefont{Romero, Flores, and
  Goldberg}}]{RFG09}
\bibinfo{author}{\bibfnamefont{M.~A.} \bibnamefont{Romero}},
  \bibinfo{author}{\bibfnamefont{F.}~\bibnamefont{Flores}}, \bibnamefont{and}
  \bibinfo{author}{\bibfnamefont{E.~C.} \bibnamefont{Goldberg}},
  \bibinfo{journal}{Phys. Rev. B} \textbf{\bibinfo{volume}{80}},
  \bibinfo{pages}{235427} (\bibinfo{year}{2009}).

\bibitem[{\citenamefont{Bajales et~al.}(2007)\citenamefont{Bajales, Ferr\'{o}n,
  and Goldberg}}]{BFG07}
\bibinfo{author}{\bibfnamefont{N.}~\bibnamefont{Bajales}},
  \bibinfo{author}{\bibfnamefont{J.}~\bibnamefont{Ferr\'{o}n}},
  \bibnamefont{and} \bibinfo{author}{\bibfnamefont{E.~C.}
  \bibnamefont{Goldberg}}, \bibinfo{journal}{Phys. Rev. B}
  \textbf{\bibinfo{volume}{76}}, \bibinfo{pages}{245431}
  (\bibinfo{year}{2007}).

\bibitem[{\citenamefont{Goldberg et~al.}(2005)\citenamefont{Goldberg, Flores,
  and Monreal}}]{GFM05}
\bibinfo{author}{\bibfnamefont{E.~C.} \bibnamefont{Goldberg}},
  \bibinfo{author}{\bibfnamefont{F.}~\bibnamefont{Flores}}, \bibnamefont{and}
  \bibinfo{author}{\bibfnamefont{R.~C.} \bibnamefont{Monreal}},
  \bibinfo{journal}{Phys. Rev. B} \textbf{\bibinfo{volume}{71}},
  \bibinfo{pages}{035112} (\bibinfo{year}{2005}).

\bibitem[{\citenamefont{Onufriev and Marston}(1996)}]{OM96}
\bibinfo{author}{\bibfnamefont{A.~V.} \bibnamefont{Onufriev}} \bibnamefont{and}
  \bibinfo{author}{\bibfnamefont{J.~B.} \bibnamefont{Marston}},
  \bibinfo{journal}{Phys. Rev. B} \textbf{\bibinfo{volume}{53}},
  \bibinfo{pages}{13340} (\bibinfo{year}{1996}).

\bibitem[{\citenamefont{Marston et~al.}(1993)\citenamefont{Marston, Andersson,
  Behringer, Cooper, DiRubio, Kimmel, and Richardson}}]{MAB93}
\bibinfo{author}{\bibfnamefont{J.~B.} \bibnamefont{Marston}},
  \bibinfo{author}{\bibfnamefont{D.~R.} \bibnamefont{Andersson}},
  \bibinfo{author}{\bibfnamefont{E.~R.} \bibnamefont{Behringer}},
  \bibinfo{author}{\bibfnamefont{B.~H.} \bibnamefont{Cooper}},
  \bibinfo{author}{\bibfnamefont{C.~A.} \bibnamefont{DiRubio}},
  \bibinfo{author}{\bibfnamefont{G.~A.} \bibnamefont{Kimmel}},
  \bibnamefont{and}
  \bibinfo{author}{\bibfnamefont{C.}~\bibnamefont{Richardson}},
  \bibinfo{journal}{Phys. Rev. B} \textbf{\bibinfo{volume}{48}},
  \bibinfo{pages}{7809} (\bibinfo{year}{1993}).

\bibitem[{\citenamefont{Clementi and Roetti}(1974)}]{CR74}
\bibinfo{author}{\bibfnamefont{E.}~\bibnamefont{Clementi}} \bibnamefont{and}
  \bibinfo{author}{\bibfnamefont{C.}~\bibnamefont{Roetti}},
  \bibinfo{journal}{Atomic. Data Nucl. Data Tables}
  \textbf{\bibinfo{volume}{14}}, \bibinfo{pages}{177} (\bibinfo{year}{1974}).

\bibitem[{\citenamefont{Gadzuk}(1967{\natexlab{a}})}]{Gadzuk67a}
\bibinfo{author}{\bibfnamefont{J.~W.} \bibnamefont{Gadzuk}},
  \bibinfo{journal}{Surface Science} \textbf{\bibinfo{volume}{6}},
  \bibinfo{pages}{133} (\bibinfo{year}{1967}{\natexlab{a}}).

\bibitem[{\citenamefont{Gadzuk}(1967{\natexlab{b}})}]{Gadzuk67b}
\bibinfo{author}{\bibfnamefont{J.~W.} \bibnamefont{Gadzuk}},
  \bibinfo{journal}{Surface Science} \textbf{\bibinfo{volume}{6}},
  \bibinfo{pages}{159} (\bibinfo{year}{1967}{\natexlab{b}}).

\bibitem[{\citenamefont{Gadzuk}(2009)}]{Gadzuk09}
\bibinfo{author}{\bibfnamefont{J.~W.} \bibnamefont{Gadzuk}},
  \bibinfo{journal}{Phys. Rev. B} \textbf{\bibinfo{volume}{79}},
  \bibinfo{pages}{073411} (\bibinfo{year}{2009}).

\bibitem[{\citenamefont{Coleman}(1984)}]{Coleman84}
\bibinfo{author}{\bibfnamefont{P.}~\bibnamefont{Coleman}},
  \bibinfo{journal}{Phys. Rev. B} \textbf{\bibinfo{volume}{29}},
  \bibinfo{pages}{3035} (\bibinfo{year}{1984}).

\bibitem[{\citenamefont{Kotliar and Ruckenstein}(1986)}]{KR86}
\bibinfo{author}{\bibfnamefont{G.}~\bibnamefont{Kotliar}} \bibnamefont{and}
  \bibinfo{author}{\bibfnamefont{A.~E.} \bibnamefont{Ruckenstein}},
  \bibinfo{journal}{Phys. Rev. Lett.} \textbf{\bibinfo{volume}{57}},
  \bibinfo{pages}{1362} (\bibinfo{year}{1986}).

\bibitem[{\citenamefont{Kadanoff and Baym}(1962)}]{KB62}
\bibinfo{author}{\bibfnamefont{L.~P.} \bibnamefont{Kadanoff}} \bibnamefont{and}
  \bibinfo{author}{\bibfnamefont{G.}~\bibnamefont{Baym}},
  \emph{\bibinfo{title}{{Quantum Statistical Mechanics}}}
  (\bibinfo{publisher}{Benjamin, New York}, \bibinfo{year}{1962}).

\bibitem[{\citenamefont{Keldysh}(1965)}]{Keldysh65}
\bibinfo{author}{\bibfnamefont{L.~V.} \bibnamefont{Keldysh}},
  \bibinfo{journal}{Sov. Phys. JETP} \textbf{\bibinfo{volume}{20}},
  \bibinfo{pages}{1018} (\bibinfo{year}{1965}).

\bibitem[{\citenamefont{Otsuki and Kuramoto}(2006)}]{OK06}
\bibinfo{author}{\bibfnamefont{J.}~\bibnamefont{Otsuki}} \bibnamefont{and}
  \bibinfo{author}{\bibfnamefont{Y.}~\bibnamefont{Kuramoto}},
  \bibinfo{journal}{J. Phys. Soc. Jpn.} \textbf{\bibinfo{volume}{75}},
  \bibinfo{pages}{064707} (\bibinfo{year}{2006}).

\bibitem[{\citenamefont{Kroha and W\"{o}lfle}(2005)}]{KW05}
\bibinfo{author}{\bibfnamefont{J.}~\bibnamefont{Kroha}} \bibnamefont{and}
  \bibinfo{author}{\bibfnamefont{P.}~\bibnamefont{W\"{o}lfle}},
  \bibinfo{journal}{J. Phys. Soc. Jpn.} \textbf{\bibinfo{volume}{74}},
  \bibinfo{pages}{16} (\bibinfo{year}{2005}).

\bibitem[{\citenamefont{Pruschke and Grewe}(1989)}]{PG89}
\bibinfo{author}{\bibfnamefont{T.}~\bibnamefont{Pruschke}} \bibnamefont{and}
  \bibinfo{author}{\bibfnamefont{N.}~\bibnamefont{Grewe}}, \bibinfo{journal}{Z.
  Phys. B} \textbf{\bibinfo{volume}{74}}, \bibinfo{pages}{439}
  (\bibinfo{year}{1989}).

\bibitem[{\citenamefont{Holm and Sch\"{o}nhammer}(1989)}]{HS89}
\bibinfo{author}{\bibfnamefont{J.}~\bibnamefont{Holm}} \bibnamefont{and}
  \bibinfo{author}{\bibfnamefont{K.}~\bibnamefont{Sch\"{o}nhammer}},
  \bibinfo{journal}{Solid State Commun.} \textbf{\bibinfo{volume}{69}},
  \bibinfo{pages}{969} (\bibinfo{year}{1989}).

\bibitem[{\citenamefont{Sakai et~al.}(1988)\citenamefont{Sakai, Motizuki, and
  Kasuya}}]{Sakai88}
\bibinfo{author}{\bibfnamefont{O.}~\bibnamefont{Sakai}},
  \bibinfo{author}{\bibfnamefont{M.}~\bibnamefont{Motizuki}}, \bibnamefont{and}
  \bibinfo{author}{\bibfnamefont{T.}~\bibnamefont{Kasuya}}, in
  \emph{\bibinfo{booktitle}{{Core level spectroscopy in condensed systems}}},
  edited by \bibinfo{editor}{\bibfnamefont{J.}~\bibnamefont{Kanamori}}
  \bibnamefont{and} \bibinfo{editor}{\bibfnamefont{A.}~\bibnamefont{Kotani}}
  (\bibinfo{publisher}{Springer Verlag}, \bibinfo{address}{Berlin},
  \bibinfo{year}{1988}), p.~\bibinfo{pages}{45}.

\bibitem[{\citenamefont{Nordlander and Tully}(1988)}]{NT88}
\bibinfo{author}{\bibfnamefont{P.}~\bibnamefont{Nordlander}} \bibnamefont{and}
  \bibinfo{author}{\bibfnamefont{J.~C.} \bibnamefont{Tully}},
  \bibinfo{journal}{Phys. Rev. Lett.} \textbf{\bibinfo{volume}{61}},
  \bibinfo{pages}{990} (\bibinfo{year}{1988}).

\bibitem[{\citenamefont{Nordlander and Tully}(1990)}]{NT90}
\bibinfo{author}{\bibfnamefont{P.}~\bibnamefont{Nordlander}} \bibnamefont{and}
  \bibinfo{author}{\bibfnamefont{J.~C.} \bibnamefont{Tully}},
  \bibinfo{journal}{Phys. Rev. B} \textbf{\bibinfo{volume}{42}},
  \bibinfo{pages}{5564} (\bibinfo{year}{1990}).

\bibitem[{\citenamefont{Borisov and Wille}(1995)}]{BW95}
\bibinfo{author}{\bibfnamefont{A.~G.} \bibnamefont{Borisov}} \bibnamefont{and}
  \bibinfo{author}{\bibfnamefont{U.}~\bibnamefont{Wille}},
  \bibinfo{journal}{Surface science} \textbf{\bibinfo{volume}{338}},
  \bibinfo{pages}{L875} (\bibinfo{year}{1995}).

\bibitem[{\citenamefont{More et~al.}(1998)\citenamefont{More, Merino, Monreal,
  Pou, and Flores}}]{MMM98}
\bibinfo{author}{\bibfnamefont{W.}~\bibnamefont{More}},
  \bibinfo{author}{\bibfnamefont{J.}~\bibnamefont{Merino}},
  \bibinfo{author}{\bibfnamefont{R.}~\bibnamefont{Monreal}},
  \bibinfo{author}{\bibfnamefont{P.}~\bibnamefont{Pou}}, \bibnamefont{and}
  \bibinfo{author}{\bibfnamefont{F.}~\bibnamefont{Flores}},
  \bibinfo{journal}{Phys. Rev. B} \textbf{\bibinfo{volume}{58}},
  \bibinfo{pages}{7385} (\bibinfo{year}{1998}).

\bibitem[{\citenamefont{Vald\'{e}s et~al.}(2005)\citenamefont{Vald\'{e}s,
  Goldberg, Blanco, and Monreal}}]{VGB05}
\bibinfo{author}{\bibfnamefont{D.}~\bibnamefont{Vald\'{e}s}},
  \bibinfo{author}{\bibfnamefont{E.~C.} \bibnamefont{Goldberg}},
  \bibinfo{author}{\bibfnamefont{J.~M.} \bibnamefont{Blanco}},
  \bibnamefont{and} \bibinfo{author}{\bibfnamefont{R.~C.}
  \bibnamefont{Monreal}}, \bibinfo{journal}{Phys. Rev. B}
  \textbf{\bibinfo{volume}{71}}, \bibinfo{pages}{245417}
  (\bibinfo{year}{2005}).

\bibitem[{\citenamefont{Marbach
  et~al.}(2012{\natexlab{a}})\citenamefont{Marbach, Bronold, and
  Fehske}}]{MBF12a}
\bibinfo{author}{\bibfnamefont{J.}~\bibnamefont{Marbach}},
  \bibinfo{author}{\bibfnamefont{F.~X.} \bibnamefont{Bronold}},
  \bibnamefont{and} \bibinfo{author}{\bibfnamefont{H.}~\bibnamefont{Fehske}},
  \bibinfo{journal}{Eur. Phys. J. D} \textbf{\bibinfo{volume}{66}},
  \bibinfo{pages}{106} (\bibinfo{year}{2012}{\natexlab{a}}).

\bibitem[{\citenamefont{Marbach
  et~al.}(2012{\natexlab{b}})\citenamefont{Marbach, Bronold, and
  Fehske}}]{MBF12b}
\bibinfo{author}{\bibfnamefont{J.}~\bibnamefont{Marbach}},
  \bibinfo{author}{\bibfnamefont{F.~X.} \bibnamefont{Bronold}},
  \bibnamefont{and} \bibinfo{author}{\bibfnamefont{H.}~\bibnamefont{Fehske}},
  \bibinfo{journal}{Phys. Rev. B} \textbf{\bibinfo{volume}{86}},
  \bibinfo{pages}{115417} (\bibinfo{year}{2012}{\natexlab{b}}).

\bibitem[{\citenamefont{Marbach et~al.}(2011)\citenamefont{Marbach, Bronold,
  and Fehske}}]{MBF11}
\bibinfo{author}{\bibfnamefont{J.}~\bibnamefont{Marbach}},
  \bibinfo{author}{\bibfnamefont{F.~X.} \bibnamefont{Bronold}},
  \bibnamefont{and} \bibinfo{author}{\bibfnamefont{H.}~\bibnamefont{Fehske}},
  \bibinfo{journal}{Phys. Rev. B} \textbf{\bibinfo{volume}{84}},
  \bibinfo{pages}{085443} (\bibinfo{year}{2011}).

\bibitem[{\citenamefont{K\"{u}rpick and Thumm}(1996)}]{KT96}
\bibinfo{author}{\bibfnamefont{P.}~\bibnamefont{K\"{u}rpick}} \bibnamefont{and}
  \bibinfo{author}{\bibfnamefont{U.}~\bibnamefont{Thumm}},
  \bibinfo{journal}{Phys. Rev. A} \textbf{\bibinfo{volume}{54}},
  \bibinfo{pages}{1487} (\bibinfo{year}{1996}).

\bibitem[{\citenamefont{Slater}(1930)}]{Slater30}
\bibinfo{author}{\bibfnamefont{J.~C.} \bibnamefont{Slater}},
  \bibinfo{journal}{Phys. Rev.} \textbf{\bibinfo{volume}{36}},
  \bibinfo{pages}{57} (\bibinfo{year}{1930}).

\bibitem[{\citenamefont{Abramowitz and Stegun}(1973)}]{AS73}
\bibinfo{editor}{\bibfnamefont{M.}~\bibnamefont{Abramowitz}} \bibnamefont{and}
  \bibinfo{editor}{\bibfnamefont{I.~A.} \bibnamefont{Stegun}}, eds.,
  \emph{\bibinfo{title}{Handbook of mathematical functions}}
  (\bibinfo{publisher}{Dover Publications, Inc.}, \bibinfo{address}{New York},
  \bibinfo{year}{1973}).

\bibitem[{\citenamefont{Langreth and Wilkins}(1972)}]{LW72}
\bibinfo{author}{\bibfnamefont{D.~C.} \bibnamefont{Langreth}} \bibnamefont{and}
  \bibinfo{author}{\bibfnamefont{J.~W.} \bibnamefont{Wilkins}},
  \bibinfo{journal}{Phys. Rev. B} \textbf{\bibinfo{volume}{6}},
  \bibinfo{pages}{3189} (\bibinfo{year}{1972}).

\bibitem[{\citenamefont{Sosolik et~al.}(2003)\citenamefont{Sosolik, Hampton,
  Lavery, Cooper, and Marston}}]{SHL03}
\bibinfo{author}{\bibfnamefont{C.~E.} \bibnamefont{Sosolik}},
  \bibinfo{author}{\bibfnamefont{J.~R.} \bibnamefont{Hampton}},
  \bibinfo{author}{\bibfnamefont{A.~C.} \bibnamefont{Lavery}},
  \bibinfo{author}{\bibfnamefont{B.~H.} \bibnamefont{Cooper}},
  \bibnamefont{and} \bibinfo{author}{\bibfnamefont{J.~B.}
  \bibnamefont{Marston}}, \bibinfo{journal}{Phys. Rev. Lett.}
  \textbf{\bibinfo{volume}{90}}, \bibinfo{pages}{013201}
  (\bibinfo{year}{2003}).

\bibitem[{\citenamefont{Plihal et~al.}(1999)\citenamefont{Plihal, Langreth, and
  Nordlander}}]{PLN99}
\bibinfo{author}{\bibfnamefont{M.}~\bibnamefont{Plihal}},
  \bibinfo{author}{\bibfnamefont{D.~C.} \bibnamefont{Langreth}},
  \bibnamefont{and}
  \bibinfo{author}{\bibfnamefont{P.}~\bibnamefont{Nordlander}},
  \bibinfo{journal}{Phys. Rev. B} \textbf{\bibinfo{volume}{59}},
  \bibinfo{pages}{13322} (\bibinfo{year}{1999}).

\bibitem[{\citenamefont{Baker~Jr.}(1958)}]{Baker58}
\bibinfo{author}{\bibfnamefont{G.~A.} \bibnamefont{Baker~Jr.}},
  \bibinfo{journal}{Phys. Rev.} \textbf{\bibinfo{volume}{109}},
  \bibinfo{pages}{2198} (\bibinfo{year}{1958}).

\bibitem[{\citenamefont{Lifshitz and Pitaevskii}(1981)}]{LL81}
\bibinfo{author}{\bibfnamefont{E.~M.} \bibnamefont{Lifshitz}} \bibnamefont{and}
  \bibinfo{author}{\bibfnamefont{L.~P.} \bibnamefont{Pitaevskii}},
  \emph{\bibinfo{title}{Physical Kinetics}} (\bibinfo{publisher}{Pergamon
  Press}, \bibinfo{address}{New York}, \bibinfo{year}{1981}).

\bibitem[{\citenamefont{Bleistein and Handelsman}(1986)}]{BH86}
\bibinfo{author}{\bibfnamefont{N.}~\bibnamefont{Bleistein}} \bibnamefont{and}
  \bibinfo{author}{\bibfnamefont{R.~A.} \bibnamefont{Handelsman}},
  \emph{\bibinfo{title}{Asymptotic expansion of integrals}}
  (\bibinfo{publisher}{Dover publications}, \bibinfo{address}{New York},
  \bibinfo{year}{1986}).

\end{thebibliography}

\end{document}